\documentclass[traditabstract]{aa}

\usepackage{natbib}
\usepackage{amssymb,amsfonts}
\usepackage{amsmath,mathtools}
\usepackage{url}
\usepackage{hyperref}
\usepackage{algorithm}
\usepackage{algorithmic}
\usepackage{xcolor}
\usepackage{ulem}
\usepackage{txfonts} 

\newcommand{\LL}{\mathcal{L}}
\newcommand{\Llam}{\mathcal{L}_\Lambda}
\newcommand{\Lloc}{\mathcal{L}_{\rm L}}
\newcommand{\NPsi}{{N_\psi}}
\newcommand{\NXi}{{N_\xi}}
\newcommand{\Mbaf}{{M_\psi}}
\newcommand{\MbafXi}{{M_\xi}}
\newcommand{\Nvar}{{N_{\rm V}}}
\newcommand{\Npix}{{N_{\rm pix}}}
\newcommand{\Nlambda}{{N_{\lambda}}}
\newcommand{\NOmega}{{N_{\vec\Omega}}}
\newcommand{\Npp}{{N_{\Lambda}}}
\newcommand{\Npx}{{N_{\chi}}}
\newcommand{\Nloc}{{N_{\rm L}}}
\newcommand{\Nit}{{N_{\rm it}}}
\newcommand{\ESE}{{\mathcal{E}}}
\newcommand{\dd}{\mathrm{d}}
\DeclareMathOperator*{\argmin}{arg\,min}

\begin{document}

\title{Novel framework for the three-dimensional \\ NLTE inverse problem}

\titlerunning{3D NLTE inversion}

\author{
Ji\v{r}\'{\i} \v{S}t\v{e}p\'an\inst{1}
\and 
Tanaus\'u del Pino Alem\'an\inst{2,3}
\and
Javier Trujillo Bueno\inst{2,3,4}
}

\authorrunning{J. \v{S}t\v{e}p\'an et al.}

\institute{
Astronomical Institute of the Academy of Sciences, Ond\v{r}ejov, Czech Republic.
\email{jiri.stepan@asu.cas.cz}
\and
Instituto de Astrof\'{\i}sica de Canarias, E-38205 La Laguna, Tenerife, Spain
\and
Departamento de Astrof\'{\i}sica, Universidad de La Laguna, E-38206 La Laguna, Tenerife, Spain
\and
Consejo Superior de Investigaciones Cient\'{\i}ficas, Spain
}

\date{Received XXXX / Accepted XXXX}

\abstract{
The inversion of spectropolarimetric observations of the solar upper atmosphere is one of the most challenging goals in solar physics. If we account for all relevant ingredients of the spectral line formation process, such as the three-dimensional (3D) radiative transfer out of local thermodynamic equilibrium (NLTE), the task becomes extremely computationally expensive. Instead of generalizing 1D methods to 3D, we have developed a new approach to the inverse problem. In our meshfree method, we do not consider the requirement of 3D\,NLTE consistency as an obstacle, but as a natural regularization with respect to the traditional pixel-by-pixel methods. This leads to more robust and less ambiguous solutions. We solve the 3D\,NLTE inverse problem as an unconstrained global minimization problem that avoids repetitive evaluations of the $\Lambda$~operator. Apart from the 3D\,NLTE consistency, the method allows us to easily include additional conditions of physical  consistency such as the zero divergence of the magnetic field. Stochastic ingredients make the method less prone to ending up within the local minima of the loss function. Our method is capable of solving the inverse problem faster by several orders
of magnitude  than by using grid-based methods. The method can provide accurate and physically consistent results if sufficient computing time is available, along with approximate solutions in the case of very complex plasma structures or limited computing time.
}


\maketitle

\section{Introduction\label{sec:intro}}

The remote sensing of magnetic fields and other physical quantities of the outer solar atmosphere is a notoriously difficult problem. The new generation of solar telescopes such as DKIST \cite[currently in the commissioning phase;][]{2020SoPh..295..172R} and EST \citep[currently in the preparatory phase;][]{2019AdSpR..63.1389J} will provide spectropolarimetric observations of unprecedented quality. In addition, there is an urgent need for\ plasma diagnostic tools to aid in their correct interpretation. However, our modeling techniques lag behind the capabilities of such new observational facilities.

While the equations governing the intensity and polarization of spectral lines in solar prominences, filaments, and the chromosphere are mostly well known, the process of forward modeling remains difficult mainly for numerical reasons: accounting for all the relevant processes requires the solution of the problem of the generation and transfer of spectral line polarization in three-dimensional (3D) geometry. Since many spectral lines of interest are formed out of local thermodynamic equilibrium (NLTE), a complicated non-linear and non-local problem needs to be solved. This is the case for a number of spectral lines of high diagnostic potential whose optical thickness can approach or exceed unity, such as H$\alpha$ at 6563\,\AA\ or the \ion{He}{i} triplet at 10830\,\AA. Under NLTE conditions, we can expect a significant impact on the part of the radiative transfer within the medium and, as we show below, the full 3D\,NLTE radiative transfer already becomes inevitable  for small optical thicknesses on the order of one.

The ultimate goal of solar spectropolarimetry is to reliably infer the plasma properties from the observed data. This so-called inverse problem is even more difficult than that of forward modeling. In principle, it is necessary to explore the space of all possible model parameters and to eliminate the models that do not agree with the observations. This is not possible in practice because such parameter space is too large and a single 3D\,NLTE model evaluation already takes up a substantial amount of computing (CPU) time. We therefore need to approach the 3D\,NLTE inverse problem (3DNIP) in a different way.

A common approach to solve the inverse problem in the photosphere, chromosphere, prominences, and filaments is to use the so-called pixel-by-pixel approach in which every column of matter behind an observed pixel is treated as independent from any other. In the case of the photosphere and chromosphere, a number of techniques have been developed over the years \citep[e.g.,][]{1992ApJ...398..375R,2000ApJ...530..977S,2015A&A...577A...7S,2019A&A...626A.102A,2019A&A...623A..74D}. For recent reviews on inversion methods, see \citet{2016LRSP...13....4D} and \citet{2017SSRv..210..109D}. A number of inversion tools have also been developed for the case of prominences and filaments \citep[e.g.,][]{2003ApJ...598L..67C,2008ApJ...683..542A,2009ASPC..415..327L}. Some algorithms go beyond the pure pixel-by-pixel approach by taking into account the spatial correlation of the data \citep[e.g.,][]{2012A&A...548A...5V,2015A&A...577A.140A,2019A&A...626A.102A} but these solutions do not take into account NLTE radiative coupling among different regions of the plasma.

In the pixel-by-pixel approach to prominences, it is challenging to go beyond models in which the physical quantities are constant along the line of sight (LOS), namely,  the so-called constant-property slab approximation because there is simply no justification for considering and constructing more refined models. This approximation is indeed very rough: if we assume the prominence properties to be constant along the LOS, we would expect the properties to be also constant along a perpendicular direction, namely, in the plane of the sky -- however, the spectra are indeed changing in the plane-of-sky and in the pixel-by-pixel inversions we are interpreting due to the changing physical quantities. This approach is clearly inconsistent. An additional problem of the oversimplified models is the existence of a number of ambiguous solutions. Spectral lines that are typically optically thin, such as \ion{He}{i}\,D$_3$ at 5877\,\AA, do not suffer so much from the effects of NLTE radiative transfer \citep{2015ASSL..415..179L}, even though as subordinate lines, they can be affected by such effects due to their coupling with optically thick transitions. However, due to their negligible optical thickness, they do not provide sufficient information on the variation of the plasma parameters along the LOS. The pixel-by-pixel approach can be successfully used but only under certain physical conditions. The observed plasma structure needs to be optically thin in all directions and it must be sufficiently homogeneous along the LOS.

To the best of our knowledge, the 3DNIP has not yet been seriously studied and it has not been even clarified if it is practically possible. The existing inversion methods solve the problem by neglecting the effects of NLTE radiative transfer between different parts of the medium in the direction perpendicular to the LOS, but this leads to the neglecting of  a crucial ingredient of spectral line formation. In this paper, we introduce a new framework for solving the 3DNIP problem that takes into account the 3D NLTE radiative transfer effects. We reformulate the problem in a meshfree manner and we consider it to be an unconstrained global minimization problem. The  method can provide both exact and approximate results depending on the available CPU time. Apart from the fact that it can be relatively easily implemented, we show that it can lead to orders-of-magnitude acceleration with respect to grid-based methods. In addition, it is less prone to end up within the local minima.

In contrast to other methods, we do not consider the requirement of NLTE coupling to be an obstacle but rather to be a very strong natural regularization. The fact that all parts of the domain are coupled by a nontrivial set of NLTE equations leads to a much more robust method than if the individual LOS were considered separately. While our method is very efficient, our focus here is more on the physical consistency rather than on computational speed.

This paper is the first in a series. It mainly focuses on the inversion problem of solar prominence and filament data, but a similar approach can be applied, after some modifications, to the case of the chromosphere. In the papers to follow, we plan to address a number of details regarding the method and its practical applications.

The structure of the current paper is as follows. In Sects.~\ref{sec:formul} and~\ref{sec:method0}, we briefly review the key ingredients of the numerical NLTE spectral synthesis and inverse problems, respectively, and we discuss the specifics of the 3D solution. We also recall the concept of sparse approximations of physical quantities and we discuss the benefits of such representations. In Sect.~\ref{sec:method1}, we recall the standard minimization procedure of the inversion algorithms and we develop a new meshfree approach to the 3D inverse problem. In order to make this possible, we reformulated the inversion problem as a global optimization in which deviation from the NLTE consistency is used as a natural regularization together with other penalizations due to different physical inconsistencies. In the resulting unconstrained minimization algorithm, the atomic state is treated as independent of the magneto-hydrodynamical (MHD) state of the atmosphere. We complete the algorithm in Sect.~\ref{sec:stoch} by using a stochastic approach to the loss function minimization in order to reduce the risk of convergence to a local minimum and to make the solution more accurate and less demanding in terms of computer memory. We demonstrate the usability of the algorithm in Sect.~\ref{sec:example}, where we apply the method to invert the thermal and magnetic structure of an academic 3D\,NLTE problem and showing that the total inversion time is by two orders of magnitude smaller than it would be when using the grid-based techniques. We summarize our results in Sect.~\ref{sec:concl}, where we also provide some future prospects. Two appendices provide technical details on the implementation of the method and on the numerical example discussed in Sect.~\ref{sec:example}.

\section{Brief overview of the forward NLTE modeling\label{sec:formul}}

\begin{figure}
\begin{center}
\includegraphics[width=\columnwidth]{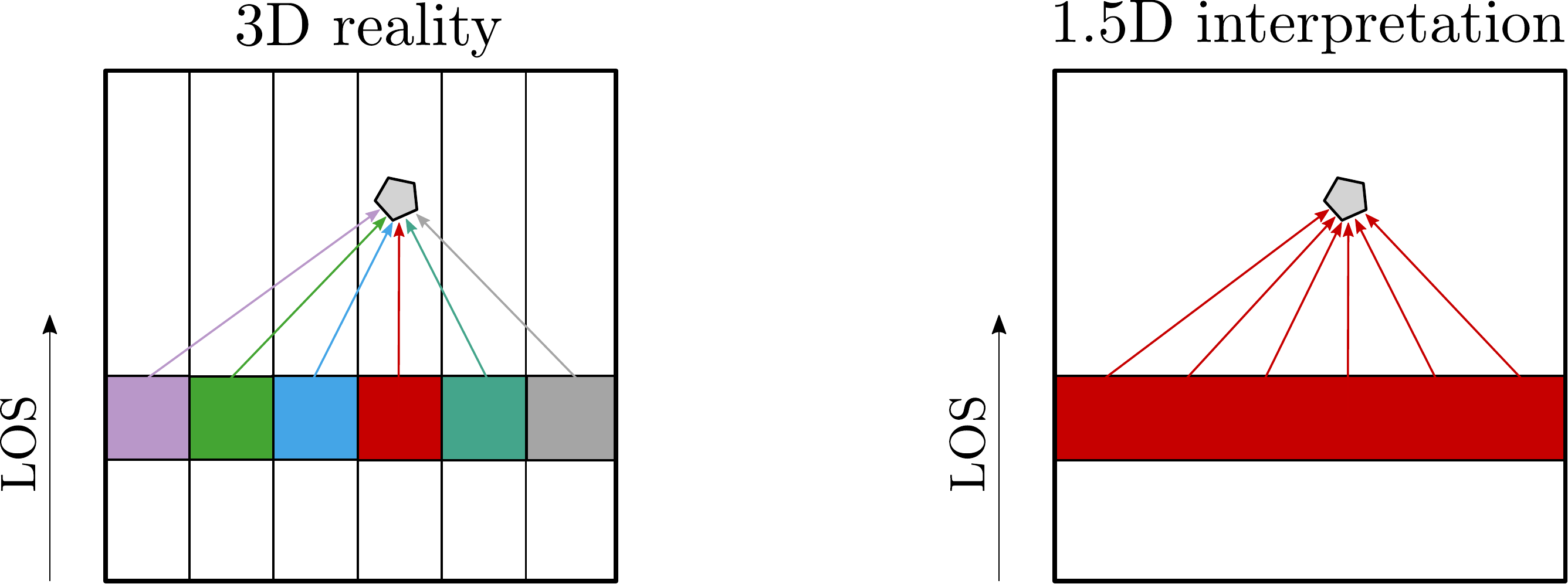}
\end{center}
\caption{Spatial coupling between lines of sight in a 3D (left) and in a 1.5D (right) model atmosphere. In 3D, the radiation field at a given point results from radiation transfer from different regions of the domain, both in the vertical and in the horizontal directions. In 1.5D, the horizontal variation is neglected. The 1.5D model can be sufficiently accurate in some cases but, in general, it often leads to serious errors in calculation of the line scattering polarization.}
\label{fig:1d3d}
\end{figure}

In this work, we develop a framework for the numerical inference of the 3D distribution of physical quantities in the solar atmosphere from spectropolarimetric data. We assume that our spectropolarimetric observation consists of a data cube with $\Npix=N^2$ spatial dimensions and $\Nlambda$ wavelengths for each of the four Stokes parameters $(I,Q,U,V)=(I_0,I_1,I_2,I_3)$, namely, the specific intensity ($I$), the two components of the linear polarization ($Q$ and $U$), and the circular polarization component ($V$). We also assume that the data are affected by Gaussian noise, but we do not consider any other instrumental degradation in this paper.

Although the main objective of this paper is to develop and describe a meshfree inference method, it is useful to compare its properties with a Cartesian grid-based method. To our knowledge, no such grid-based method has been developed before, apart from the 2.5D attempt by \v{S}t\v{e}p\'an et al. (2019; talk at the Solar Polarization Workshop~9\footnote{\url{https://www.mps.mpg.de/spw9}}). However, the numerical properties of such a method can be easily estimated. In this section, we consider (without any loss of generality) that the dimensions of the grid in the grid-based examples is $N^3$, that is, the mesh has the same spatial resolution as the data. However, take into account that in practice, it may be necessary to use a finer mesh depending on the particular spectral lines and the spatial extension of their formation region \citep[e.g.,][]{2003ASPC..288..405A}.

Given the spatial distribution of physical quantities defining the thermal and magnetic properties of the medium (such as temperature, density, magnetic field, etc.), the forward NLTE solution gives the atomic-level populations and quantum coherence that are consistent with the radiation field at every point, $\vec r$, of the medium. Hereafter, we use the term MHD-like quantities to describe this set of quantities, $\{\psi(\vec r;k)\}_{k=1}^\NPsi$, where $\NPsi$ is the number of such physical quantities. For example, if we consider the kinetic temperature of the plasma to be our first ($k=1$) MHD-like quantity, we have $\psi(\vec r;1)=T(\vec r)$.

In the standard unpolarized NLTE forward problem, the specific intensity of the radiation at any given point, propagation direction, and wavelength can be formally expressed as\footnote{For clarity, we omit the explicit mentioning of the term on the right hand side of the equation standing for the external illumination of the domain, as its inclusion does not contribute to the explanation.}
\begin{equation}
I(\vec r,\vec\Omega,\lambda)=\Lambda(\vec r,\vec\Omega,\lambda)[n(\vec{r}')]\,,
\label{eq:fslam}
\end{equation}
where $\Lambda$ is an operator which depends on the MHD-like quantities and $n(\vec {r}')$ stands for the atomic level populations within the medium. The simplicity of Eq.~(\ref{eq:fslam}) is only formal, as it represents a non-local coupling of the radiation field and the state of the matter. This equation is often referred to as the formal solution of the radiative transfer equation \citep[RTE,][]{2014HubenyMihalas}, or simply as the formal solution. It is worth noticing that, although this form of the forward problem in terms of the $\Lambda$ operator is useful in theoretical derivations, the operator is usually not explicitly constructed; it represents the process of solving the RTE.

The mean radiation field in the co-moving reference frame at every point in the medium, that is, the specific intensity integrated over the unit sphere and over the line absorption profile of spectral lines of interest is
\begin{equation}
\bar J(\vec r)=\oint \frac{\dd\vec\Omega}{4\pi} \int \dd\lambda\; I(\vec r,\vec\Omega,\lambda) \varphi(\vec r,\lambda)\,,
\label{eq:jbardef}
\end{equation}
where $\varphi(\vec r,\lambda)$ is the line's absorption profile. Likewise, we can define an averaged $\bar\Lambda$ operator that takes the above integration into account:
\begin{equation}
\bar J(\vec r)=\bar\Lambda(\vec r)[n(\vec {r}')]\,.
\label{eq:jbar}
\end{equation}

The atomic populations, the local mean radiation field, and the local MHD-like quantities are related via the equations of statistical equilibrium \citep[ESE,][]{2014HubenyMihalas}:
\begin{equation}
n(\vec r\,)=\ESE[\bar J(\vec r\,)]\,.
\label{eq:ese}
\end{equation}
Together with Eq.~(\ref{eq:jbar}), these equations make a system of coupled NLTE equations for which we must find the self-consistent distribution of atomic populations.

The general NLTE problem can only be solved numerically with iterative methods. The simplest of such methods is the so called $\Lambda$-iteration, which can be summarized in a compact form, namely: 
\begin{align}
\bar J_{k+1}(\vec r) &= \bar\Lambda(\vec r) [n_k(\vec r')]\;,  \label{eq:lamit1}    \\
n_{k+1}(\vec r) &= \ESE[\bar J_{k+1}(\vec r)]\;,    \label{eq:lamit2}
\end{align}
where the radiation field in iteration $k+1$ is calculated, at every spatial point, using the atomic populations from the previous iteration $k$ (Eq.~\ref{eq:lamit1}), and the atomic populations are, in turn, updated using this new radiation field (Eq.~\ref{eq:lamit2}). Solving Eq.~\eqref{eq:lamit1} involves the solution of the RTE at all the domain points, propagation directions, and wavelengths, and it is usually the most computationally expensive part of the NLTE problem solution (see the next section for a discussion of the numerical aspects of this step). The ESE operator $\ESE$ in Eq.~(\ref{eq:lamit2}), which depends on the MHD-like quantities, $\{\psi(\vec r;k)\}^{N_\psi}_{k=1}$, acts locally, and the computational cost of its application is rather negligible in comparison with the cost of Eq.~\eqref{eq:lamit1} for all the relevant solar physics applications.

There are numerical techniques to accelerate the solution of the NLTE problem \citep[Jacobi, Gauss-Seidel/SOR, multigrid; see, e.g.,][and references therein]{1991A&A...245..171R,1991A&A...242..290S, 1995ApJ...455..646T, 1997A&A...324..161F} that can significantly reduce the number of necessary iterations. Recently, a very promising approach to the NLTE problem acceleration has been proposed by \citet{2021arXiv211011861J} and \citet{2021arXiv211011873B}. However, the evaluation of the $\Lambda$ operator is always a key ingredient of the forward problem solution. For brevity, we use the term $\Lambda$ iteration to describe an iteration of any such numerical technique involving the evaluation of the $\Lambda$ operator, regardless of a particular acceleration scheme. 

In this work, we solve the more general problem of the generation and transfer of polarized radiation. We consider the stationary NLTE problem in complete frequency redistribution (CRD) discussed in the monograph of \citet{LL04} (the more general partial frequency distribution (PRD) problem is beyond the scope of this paper). The specific intensity is then replaced by the vector of Stokes parameters, the mean radiation field by the radiation field tensors, $\bar{J}^K_Q$, and the populations of the atomic levels by the atomic density matrix, $\rho^K_Q$. For the sake of simplicity, we only consider bound-bound transitions, but it is straightforward to generalize the method to include the bound-free and free-free transitions as well. We take into account scattering polarization, the Hanle and Zeeman effects, the macroscopic velocity gradients, and the full 3D radiative transfer, that is, all the relevant physical ingredients leading to the breaking of the symmetry of the scattering processes that critically affect the emergent polarization \citep[see][]{2015lya,2015IAUS..305..360S,2018ApJ...863..164D,2021ApJ...909..183J}.

Analogously to the MHD-like variables, we define the set of $\NXi$ atomic-like quantities $\{ \xi(\vec r;k) \}_{k=1}^\NXi$. The variables in this set are essentially the solution of the the NLTE forward problem.  There is no unique way of defining these quantities, but there are two especially useful representations. In the first, it is necessary to specify the components of the atomic density matrix, in which case the $\{\xi(\vec r;k)\}_{k=1}^\NXi$ quantities correspond to the real and imaginary parts of the density matrix elements $\rho^K_Q(\vec r; \alpha J)$ in a multi-level atom \citep[see Sect.~7.2 of][for details]{LL04}. In the second representation, we instead specify the components of the averaged radiation field tensors, $\bar J^K_Q(\vec r; \ell),$ for each of the spectral lines, $\ell$. Because the only condition that the atomic-like quantities must fulfill is to fully describe the atomic state, the two given representations are equivalent, since the radiation field tensors, $\bar J^K_Q(\vec r; \ell)$, fully determine the atomic density matrix via the ESE. In Sect.~\ref{sec:method1}, we demonstrate why it is useful to introduce the $\{ \xi(\vec r;k) \}_{k=1}^\NXi$ variables and why some representations can be more advantageous than others, depending on the specific inverse problem.

\section{Grid-based approach to the NLTE inversion}\label{sec:method0}

\subsection{Scaling of grid-based numerical forward solutions}\label{ssec:mesh}

The multi-dimensional NLTE problem of the generation and transfer of polarized radiation is typically solved on a discrete grid using the short characteristics method for the RTE \citep{1988JQSRT..39...67K}. For the sake of simplicity, we only consider Cartesian meshes in this work. As mentioned in Sect.~\ref{sec:formul}, in solar physics problems, the evaluation of the $\Lambda$ operator is typically the most computationally intensive part of the NLTE forward solution, while the evaluation time of the ESE is negligible in comparison.

In 1D problems, the computing time per $\Lambda$ iteration is proportional to the number of grid points, $N$, rays in the angular quadrature, $\NOmega$, and wavelengths, $\Nlambda$. The number of accelerated $\Lambda$ iterations of the Jacobi method (one of the most commonly used techniques) needed for convergence down to the fixed error tolerance is proportional to $N$ \citep{hackbusch85}. Consequently, the total computing time for the NLTE forward solution is on the order of $O(N^2 \NOmega \Nlambda)$.

In 3D geometry, the computing time per $\Lambda$ iteration is also proportional to the number of grid points, $N^3$. However, regarding the number of iterations, the convergence rate is determined by the distance in the units of grid steps between the most distant points. This distance is on the order of $\sqrt[3]{N^3}=N$, hence the number of accelerated $\Lambda$ iterations in 3D is usually similar to that of the 1D case. Therefore, the solution time of the 3D\,NLTE problem is typically on the order of $O(N^4\NOmega\Nlambda)$.

In order to provide a computing time example, we note that in the 3D\,NLTE code PORTA \citep{2013PORTA} the CPU time needed for evaluating the $\Lambda$ operator per grid point, propagation direction, wavelength, and Stokes parameter can be, typically, on the order of $10^{-6}$\,seconds in a common CPU. Considering a small 3D model with $N=100$ (total $100^3$ grid points), angular quadrature with 100 propagation directions, and the four Stokes parameters of a single spectral line sampled in 50 wavelengths, a single $\Lambda$ iteration would take about $5.5$~hours. With the typically expected 100 Jacobi iterations, the total computation time of the NLTE forward solution is about 550~hours. The computing time in more realistic problems of the solar chromosphere, with typical meshes of about $500^3$ grid points, becomes correspondingly larger (by two orders of magnitude) and the application of high-performance computing (HPC) becomes a requirement \citep[see][]{2015lya,2016caii}.

In order to bypass these high requirements in computational resources, it is a common practice to split the 3D problem of the $N^3$ grid into $N^2$ independent 1D problems with $N$ grid points (the so-called 1.5D approximation). When scattering polarization, the Hanle effect, and the macroscopic velocity fields do not play a significant role in the polarization of the emergent spectral line radiation, such a 1.5D solution is faster because the ensuing axial symmetry of each 1D model of every pixel or column allows us to neglect the azimuthal dependence of the radiation field. Such approach would still be approximate, as the radiative interaction between such models is neglected, but it can still provide sufficiently accurate approximations in some applications. However, when the cylindrical symmetry is broken, both 1.5D and 3D solutions have similar computing demands and, moreover, the 1.5D approach can lead to significant errors, especially when accounting for scattering polarization (see Fig.~\ref{fig:1d3d}).

\subsection{Parameterization and discrete representation of variables}\label{sssec:discrete}

In the inverse problem, we rarely try to directly find the $\{\psi(\vec r;k)\}_{k=1}^\NPsi$ values in all the grid points because, even in 1D, the number of unknowns would be too large and, moreover, such an approach leads to underdetermined problems. Instead, we can parameterize the spatial distribution of variables using a smaller number, $\Mbaf$, of parameters $\{\psi_i(k)\}_{i=1}^\Mbaf$ from which the value of each individual variable can be reconstructed at any point within the medium:
\begin{equation}
\psi(\vec r;k) = f(\vec r; \psi_1(k), \cdots, \psi_\Mbaf(k)) \,.
\label{eq:psi-f}
\end{equation}
The form of the function $f(\cdot)$ depends on the particular chosen representation of the variables. Examples of this approach are the node-based interpolation in 1D inversions \citep{1992ApJ...398..375R} and the wavelet expansion in 2D geometry proposed by \citet{2015A&A...577A.140A}.

Henceforth, we assume that each of the MHD-like quantities is represented by a set of $\Mbaf$ parameters. For the simplicity of notation, here, we chose to parameterize every MHD-like quantity using the same number $\Mbaf$ of coefficients. Therefore, the whole model, $\psi$, can be defined as an $\NPsi \Mbaf$-elements vector of these sets of parameters,
\begin{align}
\psi &= (\psi_1, \cdots, \psi_{\NPsi \Mbaf})\nonumber\\
& = ( \psi_1(1), \cdots, \psi_\Mbaf(1), \psi_1(2), \cdots, \psi_\Mbaf(\NPsi) )\,.
\label{eq:psidef}
\end{align}
However, in practice, it can be advantageous to use different numbers of parameters for different quantities. We postpone the discussion of the particular parameterization of Eq.~\eqref{eq:psi-f} to Sect.~\ref{ssec:expans}. For convenience, we assume that the elements of $\psi$ are dimensionless parameters.

It is important to emphasize that under certain conditions, we can aim to fulfill the condition $\Mbaf\ll N^3$, which leads to a significant reduction of the number of unknowns and makes this underdetermined problem a more well-defined one. Moreover, it can lead to a very significant reduction of the total CPU time needed for the inversion. We discuss and apply this condition in the following sections.

\subsection{Inverse problem definition and solution}\label{ssec:invstd}

\begin{algorithm}
  \caption{
    Inversion as the minimization of $\chi^2$ via successive solutions of self-consistent NLTE forward problems. Here, we show the gradient descent method with step $h$.
    $\psi(i)$ stands for the $i$-th iteration of the $\psi$ vector in Eq.~\eqref{eq:psidef} and $\psi_j$ stands for its $j$-th element.
    The symbol $\bar J$ is a condensed notation for the averaged radiation field tensors calculated in the grid points and $\bar\Lambda_\psi$ corresponds to the $\bar\Lambda$ operator, constructed using the $\psi$ vector.
  \label{alg:inv1}}
  \begin{algorithmic}[1]
  \STATE \textbf{Initialization}: Randomly initialize the model's MHD-like quantities $\psi(0)$.
  \STATE Given $\psi(0)$, solve the NLTE forward problem for $\bar J(0)$ via $\Lambda$ iteration.
  \STATE $i\leftarrow 0$
  \REPEAT [\textbf{Descent along the negative $\chi^2$ gradient.}]
    \STATE $i\leftarrow i+1$
    \FOR[\textbf{Loop over the model MHD-like variables.}]{$j=1$ to $\NPsi \Mbaf$}
        \STATE Calculate j-th element gradient
               $\nabla_j \chi^2=\partial \chi^2(D;\psi(i-1))/\partial\psi_j$.
               \{{\bf $\Lambda$ iteration till NLTE consistency.}\}
    \ENDFOR
    \STATE $\psi(i)\leftarrow \psi(i-1)-h \nabla\chi^2$. \{{\bf New estimate of the model parameters.}\}
    \STATE Given $\psi(i)$, solve the NLTE forward problem for $\bar J(i)$ via $\Lambda$ iteration.
  \UNTIL{$\chi^2(D;\psi(i))\approx 1$.}
  \end{algorithmic}
\end{algorithm}

Assuming that our data, $D$, consist of a square matrix of $\Npix=N^2$ pixels and $\Nlambda$ wavelengths for each of the four Stokes parameters and that they are contaminated with Gaussian noise. The inverted 3D model is parameterized by the $\psi$ vector defined in Eq.~\eqref{eq:psidef}. The measure of the goodness of the fit of the model to the data is the familiar $\chi^2$ function:
\begin{equation}
\chi^2(D;\psi) = \frac{1}{\Npix\Nlambda  } \sum_{i=1}^\Npix \sum_{j=1}^\Nlambda\sum_{k=0}^3 w_k  \frac{(I_{ijk}^{\rm M}(\psi)-I_{ijk}^{\rm O})^2}{\sigma_{ijk}^2}\,,
\label{eq:chi2}
\end{equation}
where $I_{ijk}^{\rm M}(\psi)$ is the $k$-th Stokes parameter at pixel, $i,$ and wavelength index, $j,$ calculated from the model's vector, $\psi$. Similarly, $I_{ijk}^{\rm O}$ is the corresponding observed signal, for the same $k$-th Stokes parameter at the same $i$-th pixel and $j$-th wavelength, with the noise variance $\sigma_{ijk}^2$. The $w_k$ quantities are weights for the different Stokes parameters which must fulfill the normalization condition $\sum_{k=0}^3 w_k=1$ (see Appendix \ref{app:normal}). The goal of the inversion process is to find an estimate $\hat\psi$ of the model parameters leading to the best fit to the observed data,
\begin{equation}
\hat\psi = \underset{\psi}{\argmin}\; \chi^2(D;\psi)\,.
\label{eq:psimin}
\end{equation}
Given the normalization of the signals to their noise variance, it follows that the optimal fit fulfills $\chi^2=1$. The restriction of the number of model parameters to $\Mbaf$ per quantity is an example of regularization of the solution, namely, additional constraints that cannot be solely derived from the observed data. Examples of such a regularization in the general inversion theory include the requirement on the number of non-zero parameters (i.e., sparsity regularization) or on the smoothness of the solution. Instead of Eq.~\eqref{eq:psimin}, we then solve a problem of the following form:
\begin{equation}
\hat\psi = \underset{\psi}{\argmin}\; \left[ \chi^2(D;\psi) + g(\psi) \right]\,,
\label{eq:regular}
\end{equation}
with $g(\psi)$ being the regularization condition. Regularization can help make an ill-defined problem to become well defined and it can be understood as an Occam's razor condition.

It is important to emphasize that in contrast to pixel-by-pixel inversion, in the minimization of the $\chi^2$ in Eq.~\eqref{eq:chi2}, all the pixels are inverted together because, in general, every component of the $\psi$ vector affects the model parameters in the whole computational domain (depending on the particular form of the right-hand side of Eq.~\ref{eq:psi-f}). This leads to a more robust family of methods where correlations among the MHD-like quantities in different spatial points can be naturally taken into account, such as in \citet{2015A&A...577A.140A}. Moreover, it also allows us to take full 3D radiative transfer into account, as well as other physical constraints, such as the condition of zero divergence of the magnetic field, which are practically impossible to implement in the pixel-by-pixel approaches (see below).

Algorithm~\ref{alg:inv1} shows a pseudocode for the traditional inversion process based on the minimization of the $\chi^2$ function via the gradient descent method. Even though more efficient algorithms than that of the gradient descent can be used to reduce the number of required iterations, the general structure remains the same, involving the calculation of the $\chi^2$ function gradient with respect to each model parameter (loop in lines 6--8). To calculate each gradient component $\nabla\chi^2_j$ (line 7) using a rather simple first-order method, we need to modify the $\psi_j$ variable by a small amount $\delta$, resulting in the model vector $\psi'=(\psi_1, \psi_2, \cdots, \psi_{j-1}, \psi_j+\delta, \psi_{j+1},\cdots,\psi_{\NPsi \Mbaf})$, then calculate a new self-consistent solution for $\psi'$, and calculate the new $\chi^2$ value. Thus, we obtain:
\begin{equation}
\frac{\partial \chi^2(\psi)}{\partial \psi_j}\approx\frac { \chi^2(\psi') - \chi^2(\psi)} {\delta}\,.
\label{eq:grad}
\end{equation}
Evaluating Eq.~\eqref{eq:grad} requires a NLTE forward solution, $O(\NOmega\Nlambda N^3)$, and evaluating $\chi^2(\psi')$, $O(\Nlambda N^3)$. Therefore, for $\NPsi\Mbaf$ components the calculation of the gradient (loop 6--8) scales as $O(\NPsi\Mbaf(\NOmega\Nlambda N^3+\Nlambda N^3))=O(\NPsi\Mbaf\NOmega\Nlambda N^3)$, which is completely dominated by the NLTE forward solution. Assuming that the number of inversion iterations is $\Nit$ (loop 4--11) and assuming that evaluating Eq.~(\ref{eq:grad}) requires a single $\Lambda$ iteration, the total number of $\Lambda$ iterations required for the inverse solution according to Algorithm~\ref{alg:inv1} is equal to $\Nit \NPsi \Mbaf$, implying a total scaling of $O(\Nit\NPsi\Mbaf\NOmega\Nlambda N^3)$.

We can now easily estimate the wall-clock time of a 3D inversion. Assuming, conservatively, that the number of inversion iterations is $\Nit=100$, considering that the number of MHD-like quantities would typically be on the order of $\NPsi=10$ and that the number of coefficients per quantity, $\Mbaf$, is on the order of $10^2$ (for very simple models) to $10^4$ (more refined models; see below for further details), as well as that the CPU time for each $\Lambda$ iteration is about $5.5$~hours for a model with $N^3=100^3$ (Sect.~\ref{ssec:mesh}), the total CPU time is $\Nit \NPsi \Mbaf \cdot 5.5$\,hours; namely, between 0.5 and 55 millions of CPU hours. Although enormous, such computing resources are available in today's HPC facilities. However, our experience shows that the approach sketched in Algorithm~\ref{alg:inv1} often tends to end up in a local minimum of the $\chi^2$ function and the calculation needs to be restarted. This already happens  in the 2.5D problems we tested and it can be expected that in full 3D geometry, this problem is only going to get worse. In addition, due to the possible non-uniqueness of the solution, it is nevertheless worthwhile to repeat the calculations in order to discover alternative compatible solutions. Taking into account the fact that spectropolarimetric observations with $100^2$ pixels are very coarse by today standards, the CPU time necessary for grid-based 3D inversions quickly becomes unacceptable in practice.

\section{Meshfree approach to the 3D NLTE inversion}\label{sec:method1}

In the multi-dimensional NLTE forward problem in solar-physics, the input model atmosphere is often the result of an MHD simulation, and the large-scale simulations of the outer solar atmosphere that are used in the NLTE synthesis are often computed using rectilinear Cartesian grids \citep[e.g.,][]{2011A&A...531A.154G}. Such regular discretization is advantageous in the NLTE forward problem, but the existence of a grid is a complication in the sense that the radiation transfer needs to be performed in the particular topological order of the grid nodes and, consequently, the parallelization of the NLTE solution becomes a non-trivial task.

In the inverse problem, we are not constrained by any a priori model to work with and, therefore, we are not obliged to adopt any particular spatial discretization. In this section, we introduce the basic ingredients and algorithms for a meshfree method for solving the 3DNIP. As shown below in this section, abandoning the space discretization entails a number of advantages, even though it is necessary to reformulate the standard inversion Algorithm~\ref{alg:inv1}.
One of the main properties of the new method is that it allows us to avoid 
the costly and repetitive application of the $\bar\Lambda$ operator (Sect.~\ref{sec:formul}), at
the expense of not automatically guaranteeing the full NLTE consistency 
of the problem at every step of the inversion.

\subsection{Expansion of variables into a basis}\label{ssec:expans}

In Sect.~\ref{sssec:discrete}, we show how we parameterized the spatial variation of each of the $\NPsi$ MHD-like quantities, $\{\psi(\vec r;k)\}_{k=1}^\NPsi$, with $\Mbaf$ coefficients, $\{\psi_i(k)\}_{i=1}^\Mbaf$, which, together with the function $f(\cdot)$ in Eq.~\eqref{eq:psi-f}, allow us to compute the physical quantities at any point, $\vec r$, within the medium. Henceforth, we particularize the general function $f(\cdot)$ to the linear function
\begin{equation}
\psi(\vec r;k)=\sum_{i=1}^\Mbaf \psi_i(k) \phi_i(\vec r)\,,
\label{eq:psiexp}
\end{equation}
where $\{\phi_i(\vec r)\}_{i=1}^\Mbaf$ is a certain basis of orthonormal functions in the 3D computational domain, namely,
\begin{equation}
(\phi_i,\phi_j) = \int \dd^3 r \, \phi_i(\vec r)\phi_j(\vec r)=\delta_{ij} \,,
\end{equation}
where $(\cdot,\cdot)$ stands for the inner product of the functions and the integration is done over the computational domain.

The choice of the basis functions is generally arbitrary, but different bases can have different degrees of suitability for approximating the distribution of quantities in different models and coordinate systems. A good approximation should fulfill $\Mbaf\ll N^3$, so that the number of free parameters per quantity is much smaller than the number of mesh points in an equivalent 3D grid, and should allow the approximation of the spatial variation of the quantities on both large and small scales. Such an approximation is often possible because the physical quantities are at least piece-wise continuous. However, due to the nature of the NLTE inverse problem, it is not possible to know the most suitable basis a priori and, at the same time, it is usually not possible either to find an optimal sparse subset of the basis functions as in \citet{2015A&A...577A.140A} due to the high CPU demands of such a process. Nevertheless, it is possible to improve the choice of the basis by considering heuristic methods and by using the fact that different quantities are often spatially (anti)correlated.

In this paper, we restrict ourselves to a fixed basis consisting of a set of typically smooth functions up to a certain order (such as in Appendix~\ref{ssec:bases}, where we provide an example in the 3D Cartesian coordinates). In this sense, it is important to understand that the term ``smooth model'' that we occasionally use is often adequate but generally overly restrictive; our constraint is actually on the number of basis functions, $\Mbaf$, rather than on their smoothness in any particular mathematical sense.

\subsection{Sampling the radiation field in pilot points}\label{ssec:sampling}

\begin{figure}
\begin{center}
\includegraphics[width=\columnwidth]{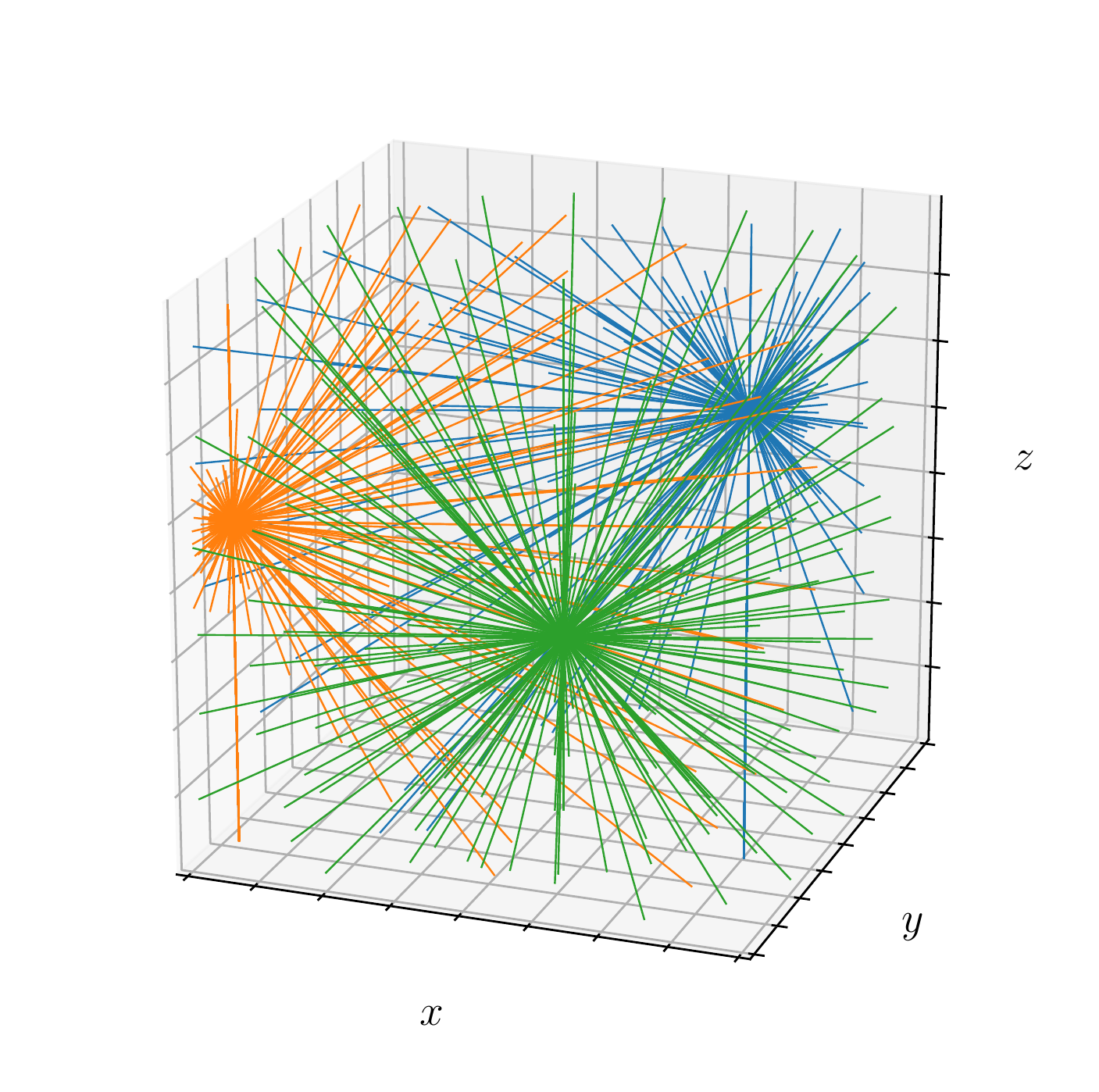}
\end{center}
\caption{Example of three pilot points and their associated long characteristics in a computational domain.}
\label{fig:lc}
\end{figure}

What makes the 3DNIP problem more difficult than other inverse problems is that, even if we find the $\psi$ correctly reproducing the observed data $D$, we still need to guarantee that the radiation and the atomic state are consistent at every point within the medium (see Sect.~\ref{sec:formul}). Similarly to the expansion of the MHD-like variables in Sect.~\ref{ssec:expans}, we can do the same with the atomic-like quantities:
\begin{equation}
\xi(\vec r;k)=\sum_{i=1}^\MbafXi \xi_i(k) \phi_i(\vec r)\,,
\label{eq:xiexp}
\end{equation}
where $\MbafXi$ is the number of basis functions. Analogously to Eq.~\eqref{eq:psidef}, we can represent the information on these variables with a $\NXi\MbafXi$ elements vector,
\begin{align}
\xi &= (\xi_1, \cdots, \xi_{\NXi \MbafXi}) \nonumber\\
&=( \xi_1(1), \cdots, \xi_\MbafXi(1), \xi_1(2), \cdots, \xi_\MbafXi(\NXi) ) \,.
\label{eq:xidef}
\end{align}

It is important to emphasize that since we assume that in the family of models of interest, it is also $\NXi\ll N^3$, we can fully determine the $\xi$ vector from a relatively small number of samples in the 3D domain. To show this, let $y(\vec r)$ be an arbitrary real function that can be expanded in the $\{\phi_i(\vec r)\}_{i=1}^M$ basis,
\begin{equation}
    y(\vec r) = \sum_{i=1}^M q_i \phi_i(\vec r),
\label{eq:ydef}
\end{equation}
where the $q_i = (y,\phi_i)$ are the expansion coefficients, and let $\{\vec {r}_j\}_{j=1}^M$ be a set of points randomly distributed in the model domain. At each of these points Eq.~\eqref{eq:ydef} becomes $y(\vec{r}_j) = \sum_{i=1}^M q_i\phi_i(\vec{r}_j)$ or, equivalently, $\vec{y} = \vec\Phi\vec{q}$, where $\vec{y}$ and $\vec{q}$ are column vectors and $\vec{\Phi}$ is a $M$ by $M$ matrix with elements $\Phi_{ji}=\phi_i(\vec r_j)$. In our approach, we restrict ourselves to the sets of random points for which the matrix $\vec\Phi$ can be inverted which are, from a practical point of view, the vast majority of cases. Due to isomorphism of $\vec q$ and $\vec y$, any random sample $\vec y$ uniquely determines the vector $\vec{q} = \vec{\Phi}^{-1}\vec{y}$.

Consequently, if we apply the same reasoning to the atomic-like variable $\xi(\vec r;k)$ sampled in $\Npp=\MbafXi$ random points\footnote{We introduce here the new variable $\Npp$ to represent the number of sampling points and we make it equal to the number of basis functions, although we go on to show in Sect.~\ref{sec:stoch} that it can be advantageous to give different values to these variables.}, these samples are sufficient to determine the $\xi(\vec r;k)$ vector in the whole 3D domain. Because $\Npp\ll N^3$, such a condensed description could lead to a significant optimization of the numerical solution in comparison to the full $\Lambda$ iteration. From now on, we refer to the random points used for sampling the atomic-like quantities as pilot points.\footnote{We note that the same term is used in the geophysical inverse problem research in a slightly different context \citep{doherty10}.}

Regardless of the particular representation of the atomic-like quantities, their sampling in the pilot points needs to be done via the calculation of the mean radiation field tensors $\bar J^K_Q(\vec r_j;\ell),$  as described in Sect.~\ref{sec:formul}. From these, the density matrix elements are easily obtained by solving the ESE.

Given an estimate of the $\psi$ and $\xi$ vectors, we can get the mean radiation field at the $\{\vec r_j\}_{j=1}^\Npp$ points by solving the RTE using the long characteristics (LC) method \citep{2014HubenyMihalas,2021ApJ...912...63D} with a suitable angular quadrature with $\NOmega$ propagation directions (see Fig.~\ref{fig:lc}). Conservatively assuming, for simplicity, that the number of discrete steps along the LC is $N$, then the time required for the calculation of the mean radiation field tensors is $O(\Npp\NOmega\Nlambda N)$, while the full $\Lambda$ iteration is of $O(\NOmega\Nlambda N^3)$. To calculate the $\chi^2$ function, we need to solve the RTE in $N^2$ pixels in order to get the emerging Stokes parameters in the whole field of view. This can also be done with LC parallel to the LOS, once per pixel, and this formal solution is then on the order of $O(N^2\Nlambda)$.

The first step towards the meshfree method discussed in this subsection is still conceptually similar to the traditional approach based on the $\Lambda$-iteration approaches: instead of determining the radiation field and atomic state (represented formally by $\vec y$ in Eq.~\ref{eq:ydef}) in all the points of a fine-spaced rectilinear grid, we may consider it solely in a smaller set of pilot points. Then the inverse of the $\vec\Phi$ matrix provides us with the model parameters $\vec q$. These can be used to reconstruct the atomic state at all points in the domain. This information can subsequently be used in the radiative transfer calculations along the long characteristics going through the whole domain. In the following section, we show that this naive generalization does not bring many benefits over the traditional $\Lambda$-iteration approach and a more radical approach is needed.

\subsection{3DNIP as an unconstrained global minimization problem}\label{ssec:unconstr}

\begin{figure}
\begin{center}
\includegraphics[width=\columnwidth]{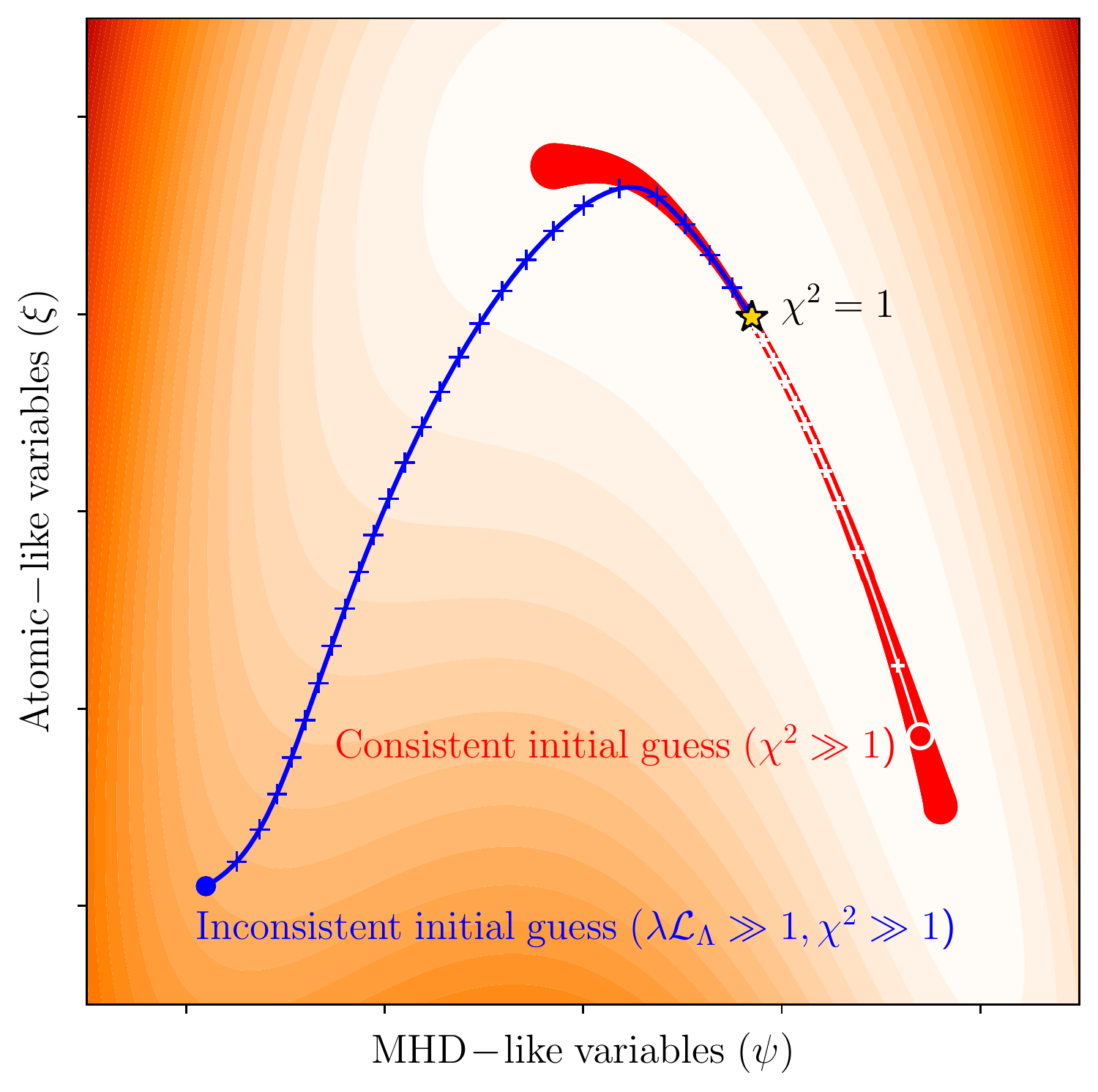}
\end{center}
\caption{Schematic visualization of the inversion process in the space spanned by the MHD-like and atomic-like variables. The background color indicates deviation from the NLTE consistency, $\Llam$, for every combination of the MHD- and atomic-like variables, with the brightest color representing the minimum value. The red line shows the common approach to the NLTE inversion as a sequence of physically consistent NLTE models of decreasing $\chi^2$ value (constrained minimization). The blue line shows an unconstrained minimization with allowed inconsistent solutions during the inversion process.}
\label{fig:unconstr}
\end{figure}

The solution of the 3DNIP in the meshfree representation of quantities could be done with minor modifications to Algorithm~\ref{alg:inv1}. The difference would be in the application of the $\bar\Lambda$ operator at lines 2, 7, and 10 of the algorithm. This operator would be evaluated just in $\Npp = \MbafXi$ pilot points instead of in the whole 3D grid. However, a closer inspection of this approach reveals that it is seriously deficient.

At line 7 in Algorithm~\ref{alg:inv1}, we modify $\psi_j$ by a small amount $\delta$, we solve the ESE in the pilot points, calculate the new radiative transfer coefficients along the LC, and then we solve the RTE in order to get the new values of the $\bar J^K_Q(\vec r_j)$ tensors. If $\delta$ is small, only a few such iterations are necessary. Finding $\bar J^K_Q(\vec r_j)$ in the pilot points is on the order of $O(\Npp \NOmega \Nlambda N)=O(\MbafXi \NOmega \Nlambda N)$. The problem is that once the $\bar J^K_Q(\vec r_j)$ in the pilot points have been found (or, equivalently, the new values of $\xi(\vec r_j,k)$, as in Sect.~\ref{ssec:sampling}), it is necessary to find the expansion coefficients $\xi_i(k)$ of Eq.~(\ref{eq:xiexp}) that are necessary for evaluating the modified $\chi^2$ function. This operation involves the $\vec\Phi^{-1}$ matrix multiplication (see the previous section). With $\NXi$ atomic-like quantities, this operation is on the order of $O(\NXi\MbafXi^2)$. Moreover, evaluating the radiative transfer coefficients for the $\chi^2$ calculation is on the order of $O(\NXi \MbafXi N^3)$ because the RTE coefficients are linear combinations of $\MbafXi$ basis functions of $\NXi$ quantities evaluated in $N^3$ points. Finally, the calculation of the $\chi^2$ is then on the order of $O(N^3\Nlambda)$.

Consequently, the loop 6--8 would be $O( \NPsi\Mbaf(\MbafXi \NOmega \Nlambda N + \NXi\MbafXi^2 + \NXi\MbafXi N^3 + N^3\Nlambda))=O( \NPsi\Mbaf(\MbafXi \NOmega \Nlambda N + \NXi\MbafXi N^3 + N^3\Nlambda))$, taking into account that $\MbafXi<N^3$. In any problem of reasonable complexity, the dominant term is $\NXi\MbafXi N^3$, hence the computing demands of the method can be written as $O( \NPsi\Mbaf \NXi\MbafXi N^3 )$.

In Sect.~\ref{ssec:invstd} we estimated the CPU time demands per iteration of the grid-based method to be $O(\NPsi\Mbaf\NOmega\Nlambda N^3)$. Since we can expect $\NXi\MbafXi>\NOmega\Nlambda$ 
in realistic models, the meshfree approach seems to bring no advantage with respect to the grid-based methods. In order to make the mesh-free method really efficient, we need to avoid the approach based on the inversion of the $\vec\Phi$ matrix and to develop a different technique.

Algorithm~\ref{alg:inv1} is an example of a constrained minimization of $\chi^2(D;\psi,\xi)$ in which we impose on $\xi$ the following constraint: the problem is always NLTE consistent. At every step of the solution, the $\xi$ variables are fully determined by $\psi$: $\xi=\xi(\psi)$, hence, we can write $\chi^2(D;\psi,\xi)=\chi^2(D;\psi)$ with the implicit condition on the NLTE consistency. The inversion process can be understood as a sequence of solutions in the space spanned by $\psi$ and $\xi$ such that, at every step, the model is NLTE consistent. In Fig.~\ref{fig:unconstr}, we represent such a process with the red curve, whose thickness is proportional to the value of $\chi^2$. In the context of Algorithm~\ref{alg:inv1}, the horizontal axis stands for the independent variables while the vertical axis stands for the dependent ones. Algorithm~\ref{alg:inv1} follows the consistency curve until a model with $\chi^2\approx 1$ is found. In order to benefit from the meshfree formulation, we need a more significant change in the inversion algorithm.

\subsubsection{NLTE consistency regularization}

In order take advantage of the meshfree formulation of the inverse problem, we need to substantially revise the traditional inversion algorithm. Let us define a vector of all model variables, $\theta$, as a union of the $\psi$ and $\xi$ vectors,
\begin{equation}
\theta=
\{\theta_1, \cdots, \theta_{\Nvar} \}
=\{\psi_1, \cdots, \psi_{\NPsi\Mbaf}, \xi_1, \cdots, \xi_{\NXi\MbafXi} \}\,,
\label{eq:thetadef}
\end{equation}
where
\begin{equation}
\Nvar=\NPsi \Mbaf + \NXi\MbafXi
\label{eq:nvar}
\end{equation}
is the total number of variables in the model, and let us formulate the inverse problem as an unconstrained minimization in which we let all the $\theta$ variables acquire any value independently of each other. Instead of the inverse problem of Eq.~\eqref{eq:psimin}, which is based on the minimization of the $\chi^2$ function, we can base the inverse problem on the minimization of a more general loss function,
\begin{equation}
\LL(D;\theta)=\chi^2(D;\theta) + \lambda \Llam(\theta)\,,
\label{eq:ll1}
\end{equation}
where $\chi^2(D;\theta)$ is the $\chi^2$ function of Eq~(\ref{eq:chi2}) but now calculated using the whole set of generally physically inconsistent parameters $\theta$ and then $\lambda$ is some positive constant, while $\Llam(\theta)$ is a non-negative differentiable function that stands for a penalty for deviation of $\theta$ from the NLTE consistency in the pilot points. In this formulation, we allow the model parameters to take unphysical values during the inversion, but deviations from self-consistency are penalized. The goal of the minimization is to find the lowest value of $\LL$, which is equivalent to finding the minimum value with $\chi^2=1$ and $\Llam=0$, corresponding to a NLTE-consistent solution that simultaneously fits the observed data. In Fig.~\ref{fig:unconstr}, such an inversion process is schematically shown with the blue curve. Starting from a physically inconsistent model that does not fit the data, we aim to find the solution that is simultaneously consistent and fits the observations. In comparing Eqs.~\eqref{eq:ll1} and \eqref{eq:regular}, we can interpret the $\lambda\Llam$ term as a regularization of the $\chi^2$ minimization problem that is due to the requirement of NLTE consistency.\footnote{This analogy needs to be taken with care because regularizations are usually conditions imposed regardless of the physical model, but we can understand this particular regularization in the context of models neglecting the NLTE transfer effects --- in this regard, the $\lambda\Llam$ term can be seen as a natural physical regularization imposing NLTE consistency.}

As mentioned above, the loss function $\Llam(\theta)$ should be differentiable and its minimum value should be equal to zero for solutions that are NLTE-consistent at every pilot point, $\vec r_j$, namely:\ 
\begin{equation}
\xi(\vec r_j;k)=\tilde\xi(\vec r_j,\theta; k)\,,
\end{equation}
where $\xi(\vec r_j,k)$ is the atomic-like $k$-th variable calculated at the pilot point, $\vec r_j$, from the current guess of the $\xi$ vector and $\tilde\xi(\vec r_j,\theta; k)$ is the same atomic-like variable, at the same pilot point, resulting from the radiative transfer via LC and for the given model parameterization, $\theta$. A suitable choice for $\Llam$ is the norm
\begin{equation}
\Llam(\theta) = \frac{1}{\Npp\NXi} \sum_{j=1}^\Npp  \sum_{k=1}^{\NXi} 
\left|  \xi(\vec r_j;k) - \tilde\xi(\vec r_j,\theta;k)  \right|^2\,.
\label{eq:llamdef}
\end{equation}

The $\lambda$ parameter in Eq.~\eqref{eq:ll1} needs to be estimated based on the particular problem and the required accuracy. If we define an acceptable error, $\Delta\xi^2$, $\lambda$ should fulfill $\lambda\Delta\xi^2 > 1$ in order to have the second term in Eq.~\eqref{eq:ll1} of at most the same order of magnitude as $\chi^2$ in the final solution. Therefore, a rough estimate of $\lambda$ can be made so that $\lambda > 1/\Delta\xi^2$. In practice, this choice needs to be made with care and with respect to each particular model and its parameterization. Moreover, the $\lambda$ parameter does not have to be constant: the decrease of its value as the solution approaches convergence can reduce the oscillatory behavior of the loss function and lead to better convergence, although we do not further discuss this aspect in this work.

In addition to the NLTE consistency, the form of the loss function in Eq.~\eqref{eq:ll1} allows us to include additional penalties, such as deviations from consistency with local physical-laws  (in the sense that they are independently fulfilled at every point, in contrast with the radiatively coupled NLTE consistency). We include these penalties in an additional term $\gamma\Lloc$ on the right-hand side of Eq.~\eqref{eq:ll1}, namely:
\begin{equation}
\LL(D;\theta)=\chi^2(D;\theta) + \lambda \Llam(\theta) + \gamma \Lloc(\theta)\,,
\label{eq:ll2}
\end{equation}
where $\gamma$ is another positive parameter of the problem (and similar considerations to those about the $\lambda$ parameter above can be made about its value). Examples of what we mean by deviation from local physical consistency are the existence of negative atomic populations, the non-zero divergence of the magnetic field vector $\nabla\cdot\vec B$, or the deviation from magnetohydrostatic equilibrium of the plasma. In the stationary models that we consider in this paper, it is also straightforward to include a penalty for deviations from the equation of continuity $\nabla\cdot(\rho\vec v)=0$ or other MHD equations.

The term $\Lloc$ in the loss function can thus be understood as an additional natural regularization due to physical consistency. It should be evaluated in a random set of $\Nloc$ points that can typically be set as  $\Nloc\gg\Npp$, while still requiring a negligible amount of CPU time in comparison to $\Llam$ because there is no need to perform radiative transfer calculations in order to evaluate it.

Likewise, $\Lloc\to 0$ as we are reaching a physically consistent solution. In Appendix~\ref{sec-app:lloc} we briefly discus the construction of the $\Lloc$ penalty. In general, we want this function to be a convex differentiable function of $\theta$ with the minimum $\Lloc=0$ for the locally consistent solutions.

In summary, with the loss function given by Eq.~\eqref{eq:ll2}, the 3DNIP problem can be formulated as
\begin{equation}
\hat \theta =\underset{\theta}{\rm arg\,min} \;\LL(D;\theta)\,,
\end{equation}
that is, to find the estimate $\hat\theta$ for the given data $D$ such that the $\LL(D;\theta)$ loss function has its minimum value equal to $1$. As we will see below, the condition $\Llam=\Lloc=0$ will be rarely satisfied exactly in practice, hence, assuming a suitable choice of $\lambda$ and $\gamma$ a typical solution fulfills $\chi^2\approx 1$, $\Llam\ll 1$, and $\Lloc\ll 1$.

\subsubsection{New inversion algorithm\label{sssec:newalg}}

\begin{algorithm}
  \caption{3DNIP solution as an unconstrained $\LL$ minimization.
           \label{alg:inv2}}
  \begin{algorithmic}[1]
  \STATE \textbf{Initialization}: Randomly initialize the model parameters $\theta(0)$.
  \STATE $i\leftarrow 0$
  \REPEAT [\textbf{Descent along the negative $\LL$ gradient.}]
    \STATE $i\leftarrow i+1$
        \FOR[\textbf{Loop over the model parameters.}]{$j=1$ to $\Nvar$}
            \STATE Calculate the gradient element
                   $\nabla_j \LL=\partial \LL(D;\theta(i-1))/\partial\theta_j$.
        \ENDFOR
    \STATE $\theta(i)\leftarrow \theta(i-1)-h \nabla\LL $. \{{\bf New estimate of the model parameters.}\}
  \UNTIL{$\LL\approx 1$ (or, explicitly, $\chi^2(D;\theta(i))\approx 1$ and $\Llam\ll 1$ and $\Lloc\ll 1$).}
  \end{algorithmic}
\end{algorithm}

We are now ready to devise a new inversion algorithm for the unconstrained NLTE inversion, namely, Algorithm~\ref{alg:inv2}. Even though it seems similar in structure to Algorithm~\ref{alg:inv1}, there are some very important differences. First, we randomly initialize all the variables $\theta$, including the atomic-like variables $\xi$. Secondly, the main loop (lines 3--8) runs until a good fit to the data is found that is simultaneously physically consistent. Finally, the gradient of the loss function in the inner loop (lines 5--7) is calculated with respect to all $\theta$ variables, that is, to both MHD-like, $\psi$, and atomic-like, $\xi$, variables.

In contrast to the step in line 7 in Algorithm~\ref{alg:inv1}, there is no need to evaluate the exact $\xi$ variables from the radiation field tensors in the pilot points because the $\xi$ variables are explicitly known at every step. A small perturbation of any of the $\psi_j$ or $\xi_j$ coefficients by $\delta$ leads to a minor modification of the RTE coefficients along the LC of the pilot points, whose calculation scales as $O(\Npp \NOmega\Nlambda N)=O(\MbafXi \NOmega\Nlambda N)$ (we recall here that we imposed $\Npp=\MbafXi$ in Sect.~\ref{ssec:sampling}). We emphasize that in this new approach the change of $\psi$ variables is not followed by the calculation of the corresponding changes of $\xi$ variables because all the variables are now independent.

The calculation of the loss function gradient consists of three terms:
\begin{equation}
\frac{\partial\LL}{\partial\theta_j}=\frac{\partial\chi^2}{\partial\theta_j}+\lambda\frac{\partial\Llam}{\partial\theta_j}+\gamma\frac{\partial\Lloc}{\partial\theta_j}\,.
\label{eq:lgrad}
\end{equation}
Regarding the scaling, the first term on the right-hand side is on the order of $O(\Nlambda N^3),$ as described in Sect.~\ref{ssec:invstd}, the second one is on the order of $O(\Npp\NOmega\Nlambda N)=O(\MbafXi\NOmega\Nlambda N)$, and the last term is on the order of $O(\Nloc)$. Even though the inner loop in Algorithm~\ref{alg:inv2} is over $\Nvar$ variables (which is larger than $\NPsi\Mbaf$ in Algorithm~\ref{alg:inv1} because we need to explicitly consider the atomic-like variables), the order of magnitude for $\Nvar$ and $\NPsi\Mbaf$ is expected to be the same in practical applications if the number of spectral lines in the problem is relatively small. The comparison with the grid-based scaling of $O(\NPsi\Mbaf\NOmega\Nlambda N^3)$ shows that if the number of iterations of both inversion methods is similar, the meshfree algorithm can be faster because all three derivatives on the right-hand side of Eq.~\eqref{eq:lgrad} are significantly faster than the NLTE-consistent derivative $\partial\chi^2(D;\psi)/\partial\psi_j$ of Algorithm~\ref{alg:inv1}.

In this work, we have assumed that the set of basis functions (and, in particular, its dimension) of the MHD-like and atomic-like quantities (Eqs.~\eqref{eq:psiexp} and \eqref{eq:xiexp}, respectively) are known. In some particular cases they can be estimated empirically but, in general, neither $\Mbaf$ nor $\MbafXi$ are known a priori and, thus, they need to be determined during the inversion process. Therefore, $\Mbaf$ must be such that the model is capable to fit the observed data and $\Nloc$ must be such that the computational domain is sufficiently finely sampled so that we can guarantee consistency of the physical quantities. This work is meant to be a discussion of the general framework of the inversion method rather than an exhaustive guideline for practical inversions and, consequently, we do not discuss   the values of $\Mbaf$ and $\Nloc$ in detail here. Instead, we simply assume that optimal values of $\Mbaf$, $\Npp=\MbafXi$, and $\Nloc$ are known and fixed.

However, what we do want to stress here is one important aspect related to the NLTE-consistency criteria. In the grid-based methods, it is possible to reach NLTE consistency with a grid of any coarseness. Indeed, this solution is only approximate due to the finite discretization of the medium: it does provide the solution in the grid points, but the values of the atomic-like variables are only approximate and dependent on the discretization \citep{1994A&A...292..599A}. Similarly, if we approximate the spatial distribution of the atomic-like variables by an expansion in $\MbafXi$ basis functions, we can find a consistent solution in $\Npp=\MbafXi$ pilot points but due to the non-linearity of the NLTE problem, the consistency is not fully guaranteed in other points unless $\MbafXi$ is so high (potentially infinite) that the solution is practically exact. In both grid-based and meshfree methods, we need to find the right compromise between accuracy and computation time.

In grid-based methods, we can estimate the error in the self-consistent solution by considering grids with different spatial resolutions \citep[see][]{1995ApJ...455..646T}. Similarly, in the meshfree method, we can increase the number of basis functions and of pilot points. When a sufficiently large value of $\Npp=\MbafXi$ is reached so that any increase leads to a negligible change of $\Llam$, it is an indication that a sufficient NLTE consistency has been achieved. We come back to this problem in Sect.~\ref{sec:stoch}.

\subsection{Parallelization and memory requirements}\label{ssec:parmem1}

The parallelization of multi-dimensional grid-based NLTE codes is difficult because the RTE needs to be solved following a particular order of grid points that depends on the radiation propagation direction. Different techniques based on domain decomposition and parallelization in the propagation directions and wavelengths have been developed in the past, such as the PORTA code \citep{2013PORTA}, which implements domain decomposition only in the vertical direction, resulting in good scaling with the number of CPUs but relatively high memory constraints, while the MULTI3D code \citep{2009ASPC..415...87L} implements a 3D domain decomposition, allowing for a larger distribution of sub-domains among CPUs but resulting in relatively worse scaling due to the need of iterating the boundary conditions of the domains.

One of the advantages of our meshfree method is that parallelization of the inversion is straightforward: each of the $\Nvar$ variables in loop 5--7 in Algorithm~\ref{alg:inv2} can be treated independently. Moreover, the RTE can also be solved independently for every LC associated with the pilot points or field of view pixels. This allows us to fully parallelize the solution with up to $\Nvar \Npp \NOmega$ (or $\Nvar N^2$ in the case of the $\chi^2$ evaluation) processes and the scaling with the number of processes will be practically linear. Given that these numbers exceed the tens of thousands in any practical application, the parallelization can be massive.

It follows, based on its parameterization, that the amount of memory needed to store the model is proportional to $\Nvar$. For models that are computable in a reasonable time with current supercomputers, $\Nvar$ would probably not exceed the order of a million significantly, which implies tens of megabytes needed to store the model. Consequently, the whole model parameterization can be stored in memory at every parallel process and no domain decomposition is ever needed.

However, from the computational point of view, it is advantageous to keep in the computer memory the LC of the pilot points and of the output pixels during the $\nabla\LL$ calculations, as well as some information on the RTE coefficients. This guarantees that fast corrections of the RTE coefficients can be calculated in the inner loop of Algorithm~\ref{alg:inv2} and the scaling discussed in Sect.~\ref{ssec:unconstr} is guaranteed. The amount of these data can become too large with an increasing number of pilot points, $\Npp$, or the improved resolution of the observation. We provide a solution for this drawback of the meshfree method in Sect.~\ref{sec:stoch}.

\subsection{Suitability of the meshfree method}\label{ssec:suitability}

\begin{figure}
\begin{center}
\includegraphics[width=0.65\columnwidth]{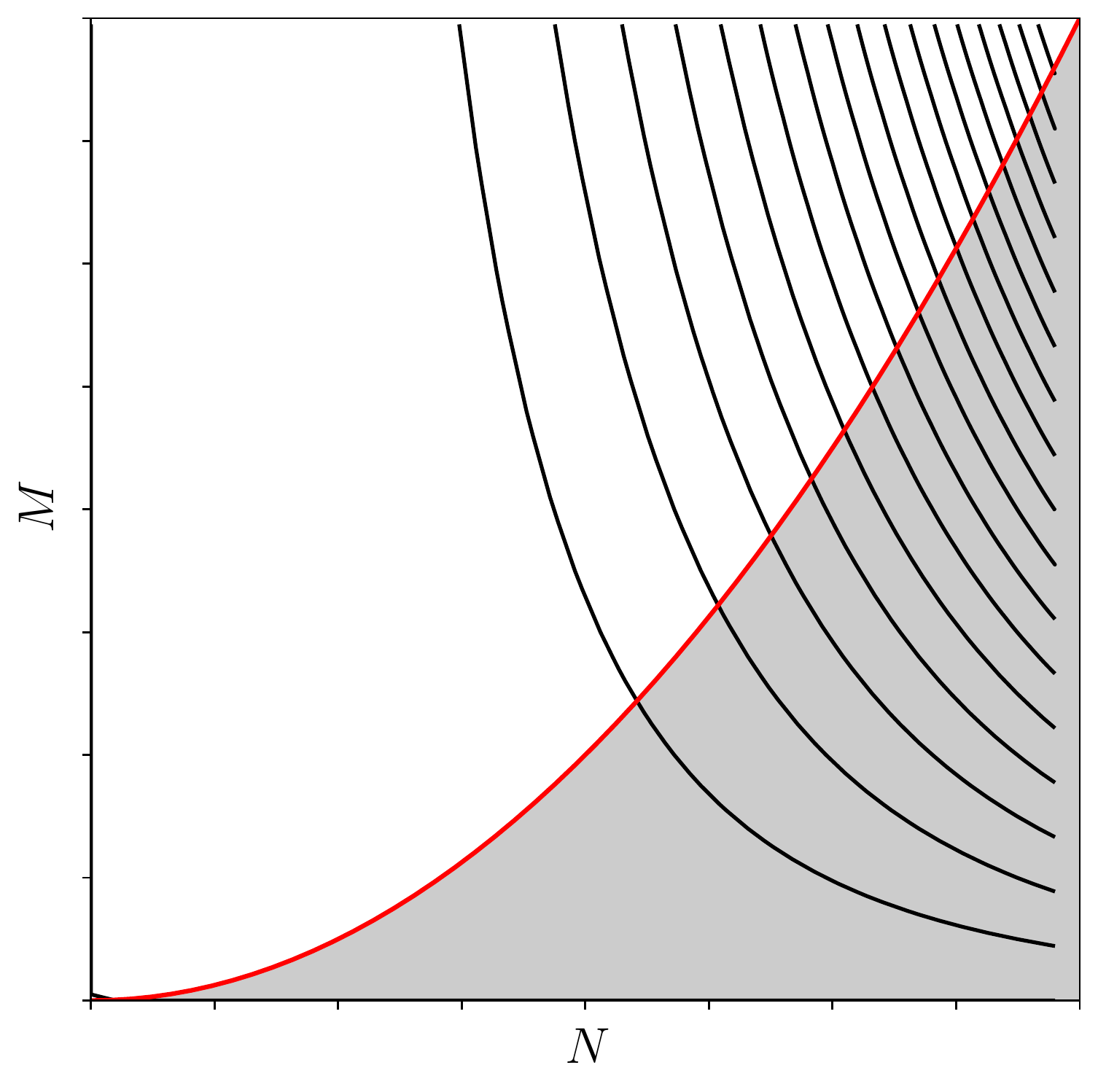}
\end{center}
\caption{Comparison between the efficiency of the meshfree and grid-based methods. The horizontal axis shows the grid resolution, while the vertical axis shows the number of coefficients per model parameter. The gray shaded area below the red curve $M=N^2$ shows the models more efficiently inverted using the meshfree method. The black curves connect models with identical required CPU time per iteration of a grid-based model. See the main text for further details.}
\label{fig:suit}
\end{figure}

Let us study the suitability of the meshfree method by comparing the required computation time of both meshfree and grid-based methods for a broad set of models. For this analysis, we assume that the number of iterations $\Nit$ needed by both methods is similar. It is difficult to analytically show that this is the general case, but our numerical experiments with 2.5D grid-based and 3D meshfree models indicate that this is likely the case in many situations. Under this assumption, the comparison of the solution with both methods is equivalent to compare just one iteration.

One straightforward condition that the meshfree method must fulfill to be more efficient is that the first and second terms on the right-hand side of Eq.~\eqref{eq:nvar} must be on a comparable order, that is, the number of $\theta$ variables should not be much larger than the number of $\psi$ variables. Given that the number of spectral lines in the model is not too great, this is usually expected to be the case.

By comparing the scaling of the calculation of each gradient component between the meshfree ($O(\Nlambda N^3 + \MbafXi\NOmega\Nlambda N + \Nloc)$, described in Sect.~\ref{sssec:newalg}, and the grid-based ($O(\NOmega\Nlambda N^3)$, described in Sect.~\ref{ssec:invstd}, methods, we can roughly estimate that the former will be faster if
\begin{equation}
\frac{1}{\NOmega} + \frac{\MbafXi}{N^2} + \frac{\Nloc}{\NOmega\Nlambda N^3} \sim \frac{\MbafXi}{N^2} < 1
\label{eq:cond}
\end{equation}
is satisfied, where we have taken into account that the first and third terms on the left-hand side of the inequality are clearly negligible.\footnote{In the 3D radiative transfer we always have $\NOmega\sim 10^2$.} In Fig.~\ref{fig:suit}, we show this resulting condition with the shaded region below the red curve in a diagram of number of basis functions, $M,$ per quantity versus the grid resolution. Here, we are assuming $\MbafXi\approx\Mbaf$, $M$ represents any of the MHD- or atomic-like quantities. The gray area where the condition in Eq.~\eqref{eq:cond} holds shows the models that are more efficiently calculated using the meshfree method.

We can thus conclude that the meshfree method is superior for models and observations with high resolution and  somewhat limited spatial variability of the parameters (smaller $M$). Likewise, the grid-based methods can be more efficient in the case of low-resolution models and observations with abrupt spatial variability of the parameters.

The black curves in Fig.~\ref{fig:suit} correspond to curves with constant $MN^3$, proportional to the CPU time per iteration of the grid-based method. The area below each of them corresponds to the models that can be inverted within a given maximum available CPU time. One important observation is that if  a limited CPU time is given (i.e., a particular black curve in the plot) and a particular model or observation resolution, $N$, then it is always possible to find a meshfree solution for certain values of  $M$ below the red curve. Given the parabolic shape of the red curve, it follows that grid-based methods are increasingly less efficient compared to the meshfree method as the model or observation resolution increases.

\section{Stochastic inversion}\label{sec:stoch}

In the previous section, we introduce several concepts leading to the formulation of a meshfree method to solve the 3DNIP. Even though the method appears promising, there are some problems that need to be solved to achieve a truly efficient algorithm, such as the fact that with $\Npp=\MbafXi$ fixed pilot points, we cannot fully guarantee the NLTE consistency in the whole computational domain or the memory allocation of a large number of LC. Additionally, in this section, we discuss the problem of local minima for $\LL$ and the artifacts introduced by the use of fixed angular quadratures in 3D.

\subsection{Reshuffling of the pilot points and pixels:  The stochastic algorithm}

One of the issues of the meshfree method, as introduced in Sect.~\ref{sec:method1} and mentioned in Sect.~\ref{sssec:newalg}, is the impossibility of ensuring the NLTE consistency in the whole domain by testing this consistency in just $\MbafXi\ll N^3$ random pilot points. One possible solution to this problem could be (after the solution has converged) to change the set of pilot points and to check whether the solution is still NLTE consistent. While this approach would result in a slower method (we would be increasing the number of required iterations without changing the cost per iteration), it directs us toward a different approach: we can generate a new set of pilot points after every $n<\Nit$ iterations, even before the convergence is reached for a given set of pilot points. In particular, we can use a small number of pilot points, $\Npp\ll\MbafXi$, which are reshuffled after every $n$ iterations (with a lower limit of $n=1$). This allows us to eventually sample the whole 3D domain. Of course, changing the set of pilot points before reaching convergence leads to a much greater final number of iterations, but these are also much faster than in the standard algorithm because of the condition $\Npp\ll\MbafXi$.

With this approach, estimating the $\Llam$ gradient goes to the order of $O(\Npp\NOmega\Nlambda N)$, which much smaller than $O(\MbafXi\NOmega\Nlambda N)$. The most time-consuming part of the $\LL$ gradient estimation would now be the $\chi^2$ function gradient, which is still on the order of $O(\Nlambda N^3)$. We would have a very fast although inaccurate estimation of the $\Llam$ gradient, but an accurate and very slow estimation of the $\chi^2$ gradient. We can thus balance the computing time of these quantities by applying a similar strategy to the calculation of the $\chi^2$ gradient, namely, instead of evaluating $\chi^2$ in $N^2$ pixels, we randomly generate $\Npx<N^2$ pixels, to be reshuffled after every iteration, in which we calculate an approximation of this quantity. The balancing of the scaling of the $\Llam$ and $\chi^2$ gradients can be done by taking $\Npp\NOmega\approx\Npx$.
Naturally, it is also possible to decrease $\Nloc$ and to apply the same reshuffling strategy to the estimation of the $\Lloc$ gradient.

This procedure can only provide an approximation to the gradient $\nabla\LL$, but every iteration becomes significantly faster than before. This approach is used in the so-called stochastic gradient descent (SGD) class of methods and it provides several benefits over the traditional gradient descent (see below). The approximate loss function now takes the form:
\begin{eqnarray}
\LL(D;\theta;\Npx,\Npp,\Nloc)=\nonumber\\
\chi^2(D;\theta;\Npx)+\lambda \Llam(\theta;\Npp) + \gamma \Lloc(\theta;\Nloc)\,.
\label{eq:llapprox}
\end{eqnarray}
The expression for $\chi^2$ is identical to that in Eq.~\eqref{eq:chi2}, except for $\Npix$ being replaced by $\Npx$. The general structure of Algorithm~\ref{alg:inv2} remains valid, but the loss function at line 6 is replaced by that of Eq.~\eqref{eq:llapprox}. Moreover, the convergence criteria at line 9 must be modified in such a way that the convergence is guaranteed not only for a single sample of $\Npp$ and $\Npx$ points, but for the whole domain. To this end, the stopping criteria should not only involve the current value of the loss function components and of their gradients, but also their values over time. We leave the discussion of this technical issue for later works.

One last change is necessary at line 8 of Algorithm~\ref{alg:inv2}, where new values of the model parameters are calculated using the estimation of the gradient. A number of SGD algorithms have been developed in the recent years and  led to much better convergence rates than the naive approach outlined in Algorithm~\ref{alg:inv2}. In our calculations, we have used the ADAM algorithm \citep{2014arXiv1412.6980K}, which uses running averages of the gradient and of its second moment in order to estimate the new values of the problem variables. We have found that it provides good convergence results for the test presented in this work (see Sect.~\ref{sec:example} for details).

In the stochastic approach outlined in this section, we replaced the small number of computationally intensive iterations from the method described in Sect.~\ref{sec:method1} by a large number of very fast iterations. With a sufficient number of iterations, we can guarantee that the solution is consistent in the whole domain and not just in $\MbafXi$ fixed points. As we show in the example below (Sect.~\ref{sec:example}), the number of stochastic iterations exceeds the thousands or tens of thousands, hence, the computational domain can be effectively sampled with a small number of $\Npp$, $\Npx$, and $\Nloc$ points.

We note that even though we are testing the NLTE consistency in just $\Npp$ pilot points during each iteration, we are also effectively probing the model along the LC themselves (cf. Fig.~\ref{fig:lc}). Due to the non-local character of the radiative transfer problem, an error accumulated along the LC will produce an error at the pilot points, in contrast to the loss function of local quantities, $\Lloc$.

Finally, the convergence analysis in the SGD is more difficult and, thus, we leave its discussion to future publications. In this work, we restrict ourselves to an empirical convergence test in Sect.~\ref{sec:example} which shows that the method can converge very quickly. The numbers $\Npp$, $\Npx$, and $\Nloc$ determine how noisy the gradient and the convergence are, with higher values leading to a better convergence rate with a less noisy gradient at the cost of larger CPU time and memory requirements per iteration.

\subsection{Convexity, local minima, and an analogy with deep learning}

The NLTE problem is strongly non-linear. We are not aware of any existing rigorous mathematical analysis of the equations for multilevel systems, but it is very likely that the $\chi^2$ and $\LL$ functions are also non-convex; hence, there are a number of local minima in which the inversion algorithm can end up. In fact, our numerical experiments with deterministic inversion algorithms show that it is not a rare occurrence.

The use of a SGD method helps to partially solve this problem because unlike deterministic methods, the stochastic crawling through the parameter space in the SGD method is not slated to remain in a local minimum. This is simply due to the fact that the exact shape of the loss function landscape (see also the background color in Fig.~\ref{fig:unconstr}) is unknown and its estimation changes between iterations; hence, the local minima may be passed through, in contrast to what happens in the standard gradient descent method with a more accurate and unchanging estimation of the loss function.

Recently, SGD methods have become important in the context of machine learning techniques. While our method is not based on these methods, an analogy can be made with the training of deep neural networks: as in our method, deep learning can be understood as a global optimization process with a substantial number of parameters. The network training proceeds by feeding a large example data set and evaluating the gradient of the loss function. In practice, it is unfeasible to use the whole database of examples in every iteration of the training. Instead, the network is fed with a relatively small number of randomly chosen examples and the loss function gradient is only approximately calculated (so-called mini-batch training). The use of a small number of random pilot points per iteration in our inversion method can be seen as analogous to the mini-batch training, while the calculation of the loss function gradient in all domain points would correspond to using the whole database of examples in every iteration of the training.

\subsection{Angular quadratures artifacts}\label{ssec:quad}

\begin{figure}
\begin{center}
\includegraphics[width=0.65\columnwidth]{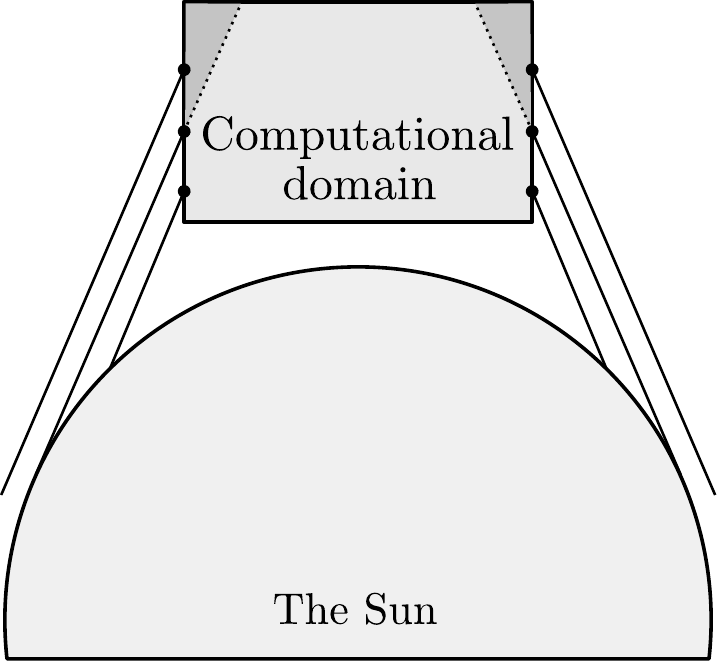}
\end{center}
\caption{Geometry of the illumination of the computational domain by the underlying solar surface using an angular quadrature. If the angular quadrature is identical at every point (see the inclined lines), artificial sharp jumps in the illumination at the boundary, which do not actually exist, will appear in the model. Such artifacts could make a full 3D inversion nearly impossible. The solution to this problem is to use a randomized quadrature at every pilot point.}
\label{fig:quad}
\end{figure}

The accurate calculation of the $\bar J^K_Q$ tensors in the pilot points requires us to solve the RTE along the LC in the directions of a sufficiently accurate angular quadrature. For a discussion of optimal quadratures for the transfer of polarized radiation, see \cite{2020quad1} and \cite{2021quad2}.

Under the physical conditions in the solar atmosphere, the number of propagation directions of a good quadrature is typically on the order of $\NOmega\sim 10^2$. As with the truncation error in the grid-based methods, the angular discretization necessarily leads to some degree of numerical error, but if an appropriate quadrature is chosen, this error can be negligible. However, a problem arises with the interaction between the quadrature and the boundary conditions of a 3D domain in the case of plasma structures embedded in the solar corona, that is, in structures such as prominences and filaments. Given a fixed quadrature and the illumination from the underlying solar surface, there will be a sharp and artificial jump in the boundary conditions, as demonstrated in Fig.~\ref{fig:quad}.

An easy solution to this problem is possible within the framework of the stochastic approach: in the same way that the pilot points are randomized in every iteration, the orientation of the quadrature rays can also be randomized in each of these points. As a result, there are no preferable directions in the radiation transfer and artifacts similar to that in Fig.~\ref{fig:quad} cannot appear. For obvious reasons, this randomization of the angular quadrature orientation cannot be implemented in the grid-based methods relying on the short-characteristics method.

In our 2.5D grid-based experiment we have found that these artifacts can actually appear in practice and they significantly complicate the solution. As a byproduct of the stochastic approach, we have the means to solve another critical problem that would actually make the full 3D inversion practically impossible.

\subsection{Parallelization and memory requirements}\label{ssec:parmem2}

The parallelization of the stochastic method is as straightforward as in the deterministic method described in Sect.~\ref{sec:method1}. Given that in every iteration we need to solve the radiative transfer along $\Nvar(\NOmega\Npp+\Npx)$ independent long characteristics. Since, in practice, we always have $\Nvar,\,\Npix>10^2$ and $\NOmega\approx 10^2$, and $\Npp$ is such that, typically, $\NOmega\Npp\approx\Npx\gtrsim 10^2$, it is clear that the method can be massively parallelized and that the scaling with the number of CPUs should be practically linear.

One of the most significant benefits of the stochastic method over the deterministic one is its random access memory requirement. As discussed in Sect.~\ref{ssec:parmem1}, it is necessary to store the data of the LC during the $\nabla\LL$ estimation, and increasing $\MbafXi$ and $N$ could dramatically increase the memory requirements to the point of this becoming a limiting factor in the method. However, in the stochastic method $\Npp\ll\MbafXi$ and $\Npx\ll N^2$ and thus the issue is non-existent. Moreover, we can always choose $\Npp$ and $\Npx$ such that the method fits any memory constraint.

\section{Example application}\label{sec:example}

\begin{figure}
\begin{center}
\includegraphics[width=\columnwidth]{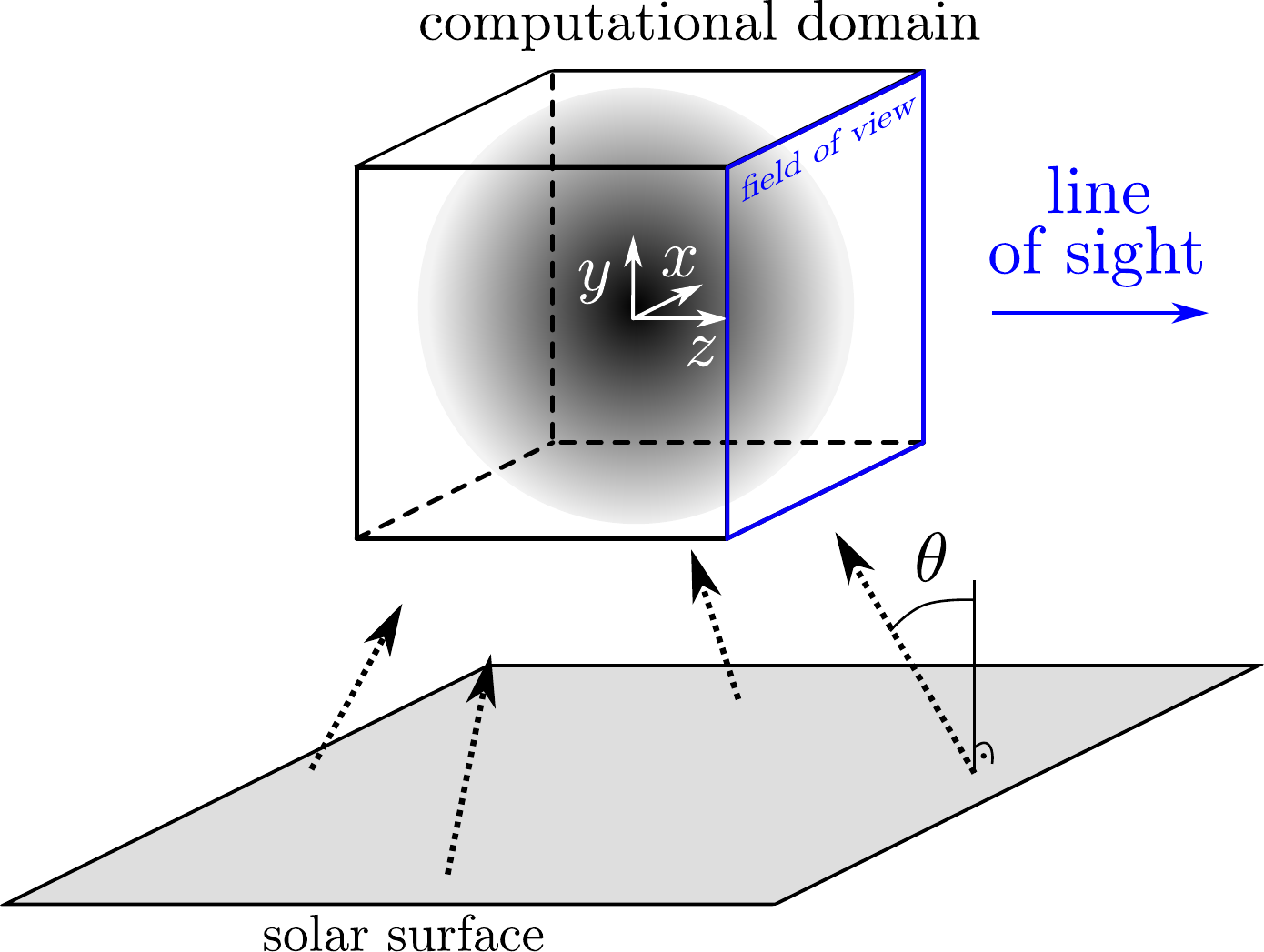}
\end{center}
\caption{Geometry of the model and the observation. The computational domain (black cube) is illuminated by the underlying Sun (gray surface). The limb darkening of the incident continuum radiation depends on the $\theta$ heliocentric angle (see the text for details). The LOS, which is in the direction of the $z$ axis, and the local solar vertical direction ($y$ axis) form a 90$^\circ$ angle. In this academic example, the solar surface is approximated by an infinite plane.}
\label{fig:geom}
\end{figure}

\begin{figure*}
\begin{center}
\includegraphics[width=0.7\columnwidth]{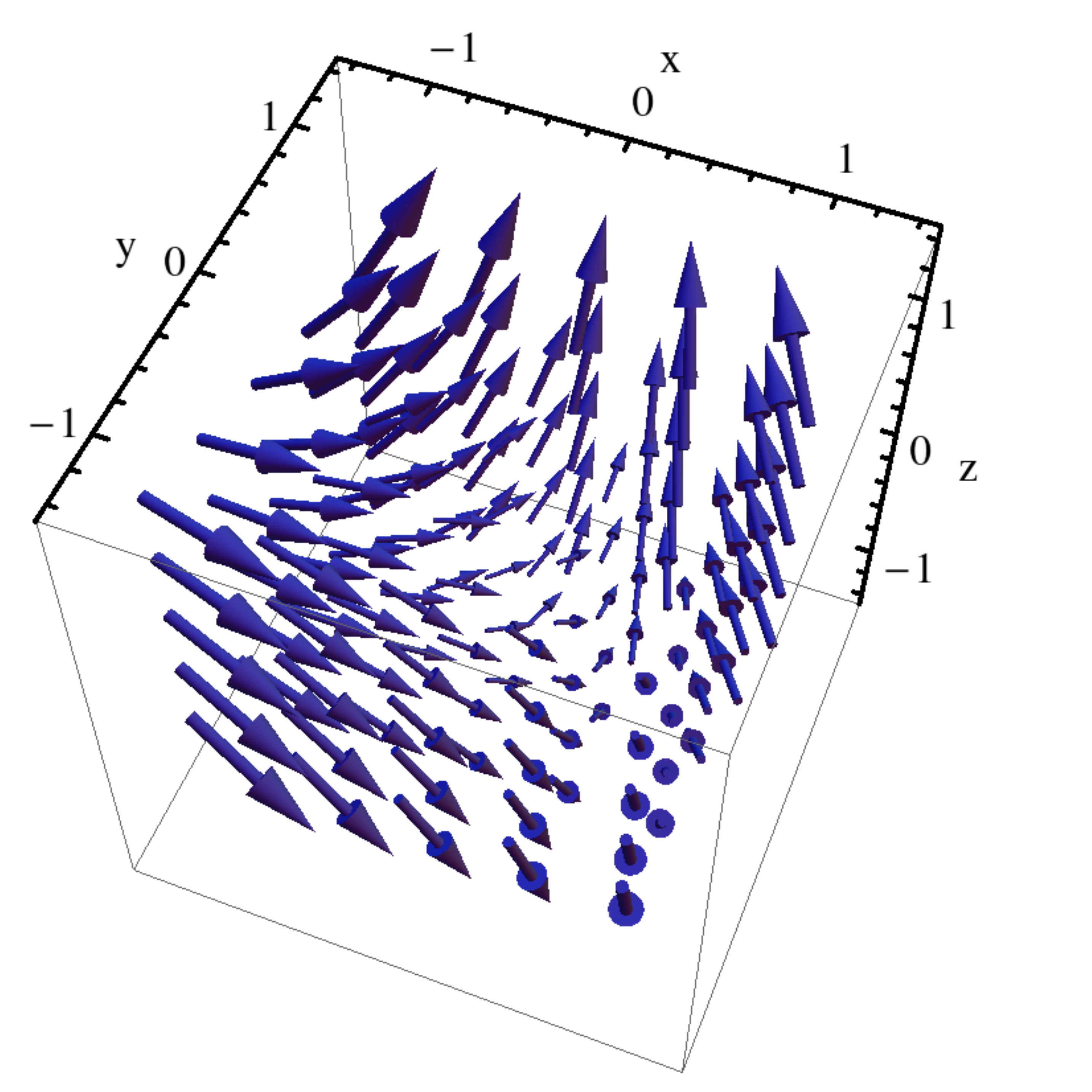}
\includegraphics[width=0.7\columnwidth]{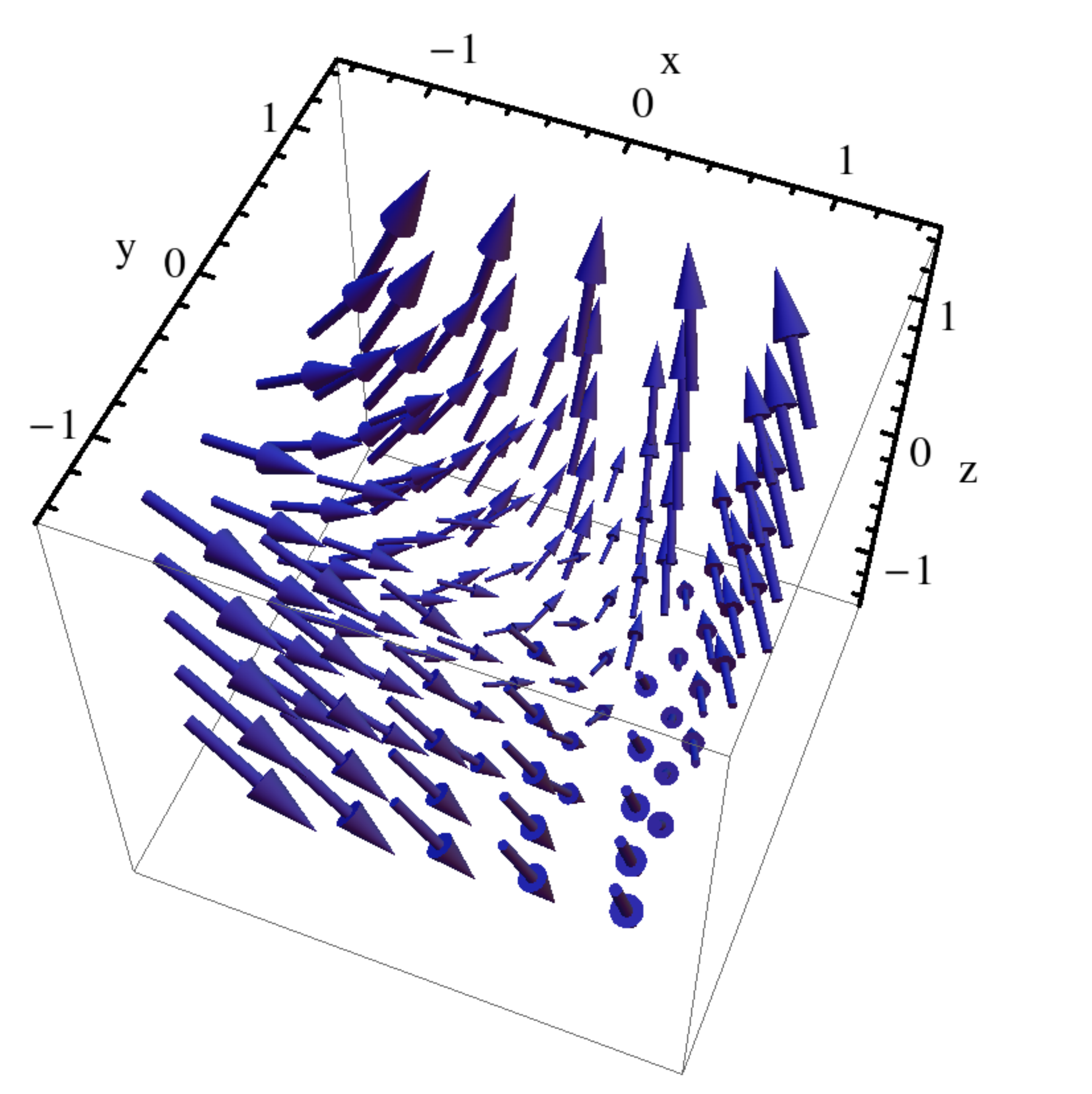}
\end{center}
\caption{Magnetic field vector in the synthetic (left) and inverted (right) models.}
\label{fig:mg}
\end{figure*}

\begin{figure*}
\begin{center}
\includegraphics[width=\textwidth]{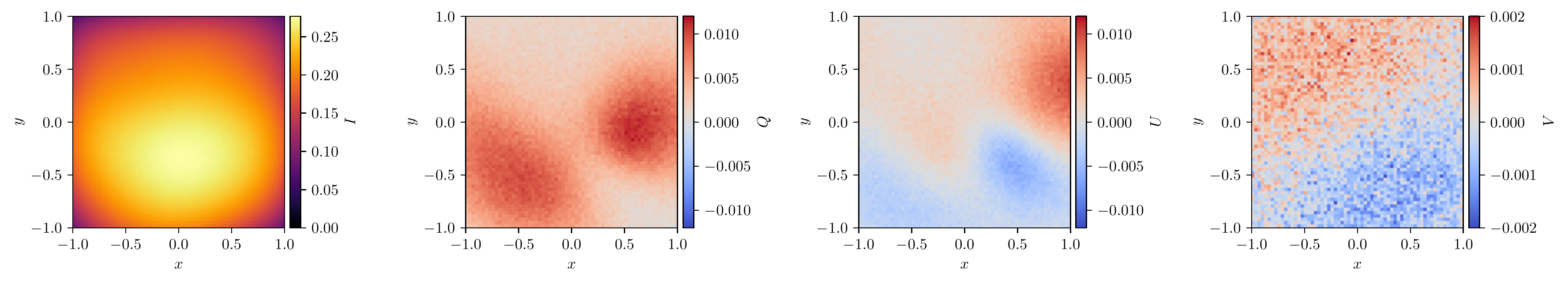} \\
\includegraphics[width=\textwidth]{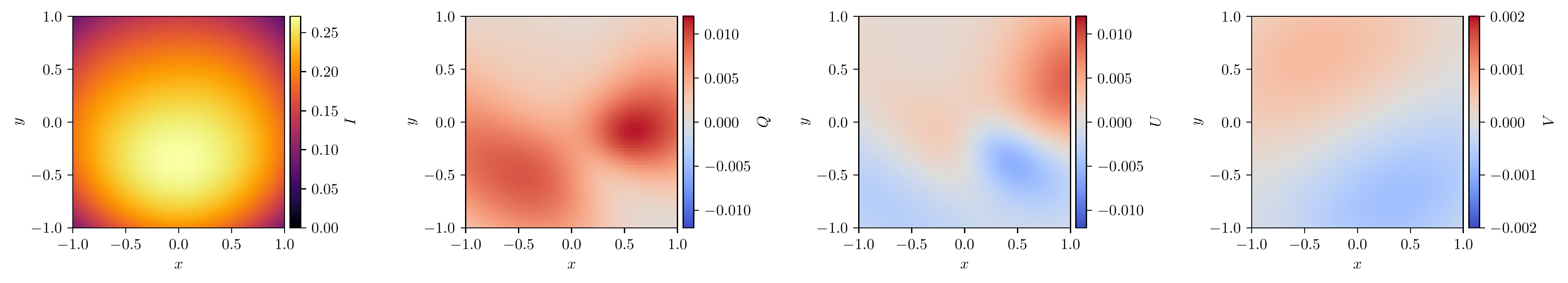}
\end{center}
\caption{
Synthetic observation (top row) and emergent radiation from the inverted model (bottom row). From left to right: $I$, $Q$, $U$ in the line center, and $V$ at around the wavelength $\lambda\approx 1.4$. The Stokes parameters are in the disk-center intensity units, $I(\mu=1)$. The positive $Q$ direction is parallel to the solar limb (parallel to the $x$ axis, cf. Fig.~\ref{fig:geom}). The synthetic signal is contaminated with Gaussian noise with $\sigma=4\cdot 10^{-4}$, in the disk-center intensity units.
}
\label{fig:obs}
\end{figure*}

\begin{figure*}
\begin{center}
\includegraphics[width=\textwidth]{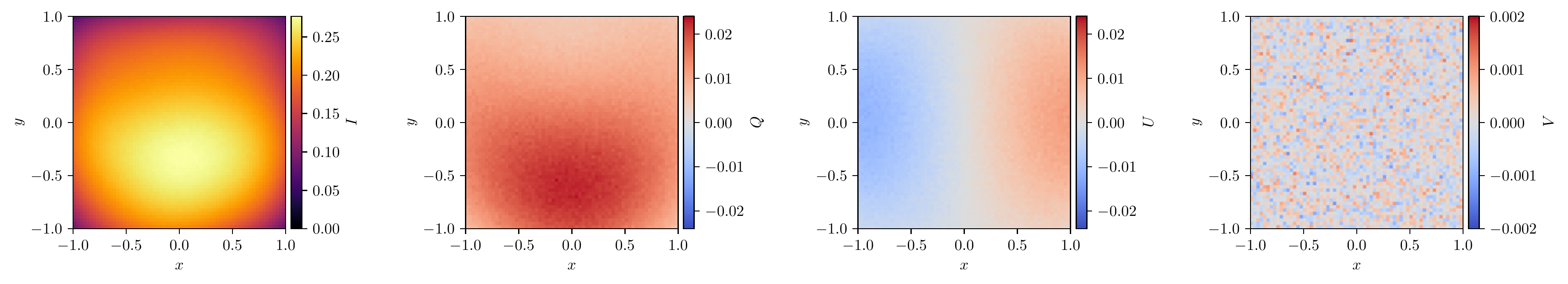}
\end{center}
\caption{Emergent radiation from the same model as in the top row in Fig.~\ref{fig:obs}, but after switching off the magnetic field. This figure demonstrates that even though the maximum optical thickness of the model is only around $\tau\approx 1$, the symmetry breaking effects due to 3D radiative transfer play a significant role and the Stokes~$U$ signal is far from zero. Neglecting 3D radiative transfer could lead to serious errors in the interpretation of the observations.}
\label{fig:obsb0}
\end{figure*}

\begin{figure*}
\begin{center}
\includegraphics[width=0.7\textwidth]{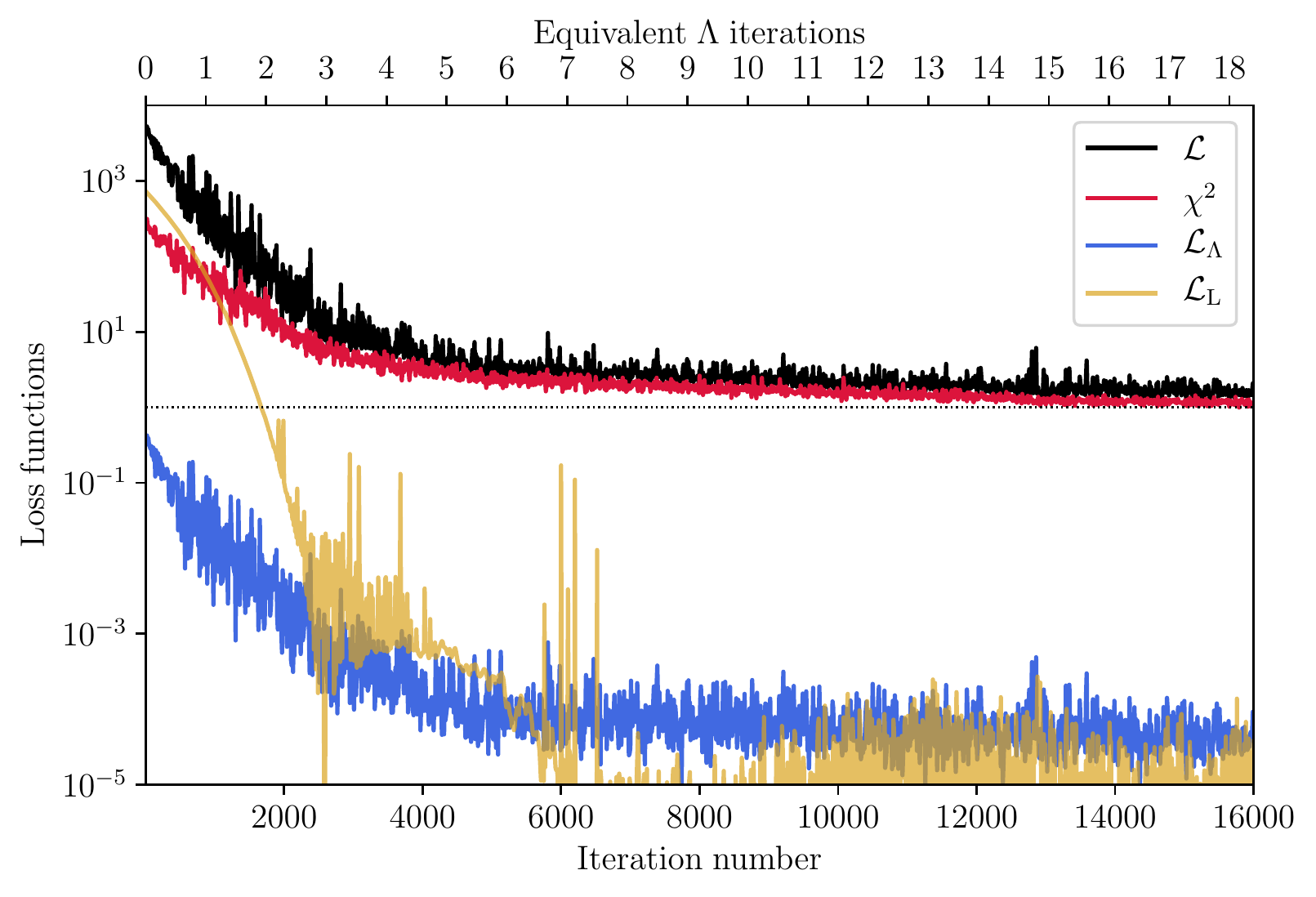}
\end{center}
\caption{Convergence of the stochastic meshfree method. The individual curves correspond to the respective terms in Eq.~(\ref{eq:llapprox}) shown in the legend. The bottom horizontal axis shows the number of stochastic iterations, while the top horizontal axis shows the CPU time in units of full $\Lambda$-iterations in a $N^3=64^3$ grid-based model.}
\label{fig:conv}
\end{figure*}

\begin{figure*}
\begin{center}
\includegraphics[width=\textwidth]{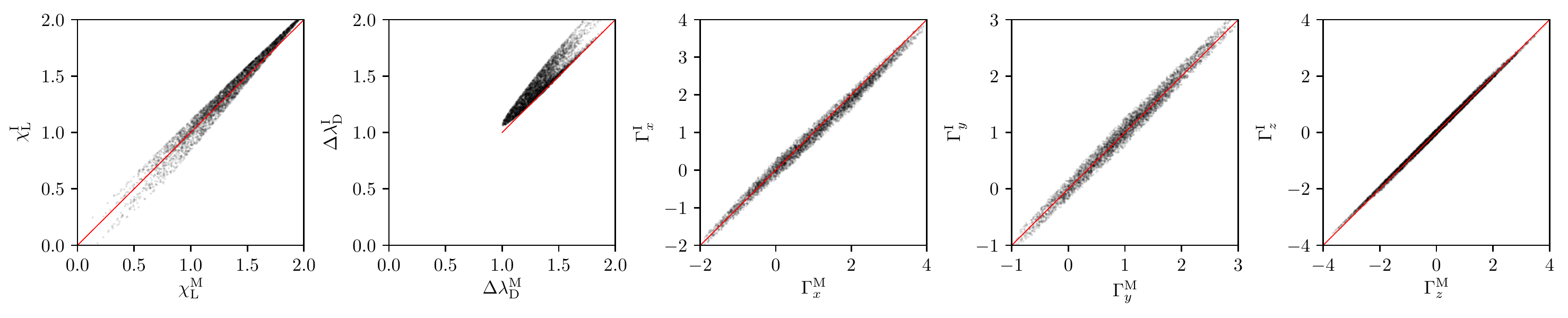}
\end{center}
\caption{Scatter plots of the true model (horizontal axes) and inverted (vertical axes) model quantities in 5\,000 randomly distributed points within the domain. From left to right: Line opacity, line Doppler width, and $x$-, $y$-, and $z$-components of the magnetic field vector.}
\label{fig:err}
\end{figure*}

\begin{figure*}
\begin{center}
\includegraphics[width=\textwidth]{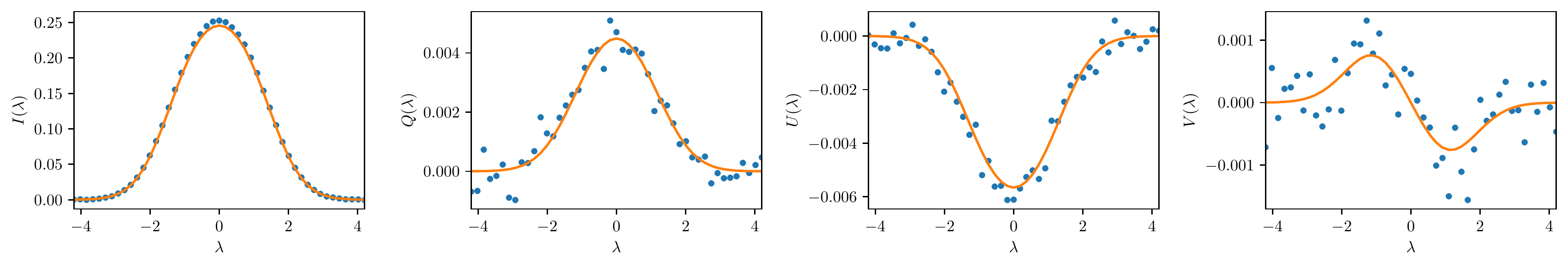}
\end{center}
\caption{Example of the inversion fit (orange curve) to the observed (blue dots) Stokes profiles at the spatial point $(x,y)=(0.75,-0.25)$}
\label{fig:profs}
\end{figure*}

In this section, we apply the previously developed stochastic meshfree method to invert the physical properties of a solar prominence-like 3D structure. For the sake of simplicity in the demonstration of the method, and without loss of generality, we use a dimensionless academic model. The chosen model is intentionally very simple and its purpose is to test the inversion algorithm.

\subsection{Model definition and synthetic observations}

The computational domain is a 3D cube with dimensions $[-1,1]^3$ suspended above the solar surface and observed along an LOS perpendicular to one of its faces (see Fig.~\ref{fig:geom}). In this example, we assume that the scattering geometry is fully known a priori during the inversion process and that we can expect the observed plasma to be completely confined within the cubic box. Needless to say, none of these assumptions would be generally satisfied in an actual observation. The boundary conditions for the illumination are chosen such that they resemble the irradiation from the underlying solar photosphere. This unpolarized spectrally flat radiation is limb-darkened according to the rule:
\begin{equation}
I(\mu)=
\begin{dcases*}
1-\tfrac 12 (1-\mu^2) & for $\mu>0$ \\
0 & for $\mu\le 0$
\end{dcases*}\;,
\label{eq:clv}
\end{equation}
where $\mu=\cos\theta$, with $\theta$ the heliocentric angle (i.e., the angle between the propagation direction and the normal to the solar surface).

At this point, we are considering an academic problem of a normal Zeeman triplet susceptible to the Hanle and Zeeman effects. Further details on the spectral line are given in Appendix \ref{ssec:app-atom}. For the sake of simplicity, we consider a spherically symmetric plasma distribution with the line opacity decreasing and the line Doppler width increasing with the distance from the domain center. This model qualitatively represents cold prominence plasma embedded in a hot surrounding corona. Details of the particular parameterization can be found in Appendix \ref{sec:app-plasma}, together with the simple, albeit non-trivial, configuration of the magnetic field vector shown in the left panel of Fig.~\ref{fig:mg}. For additional details on the model, see Appendix~\ref{sec:app-model}.

Assuming that our observation consists of the four Stokes profiles, sampled in $\Nlambda=47$ wavelengths, for a $\Npix=64\times 64$ field of view, we synthesized the observation with the PORTA code in an atmospheric model with a spatial grid of $N^3=64^3$ points. In order to mimic more realistic observations, we added random Gaussian noise with a standard deviation of $\sigma=4\cdot 10^{-4}$ in units of the disk center intensity. Given the maximum observed intensity, the noise-to-signal ratio is always larger than $10^{-3}$ in the observations. The synthetic observations are shown in the top panel of Fig.~\ref{fig:obs}.

In Fig.~\ref{fig:obsb0}, we show the same synthetic data, calculated in the same model atmosphere but without any magnetic field. While the circular polarization (Stokes~$V$) is obviously zero and the intensity is practically unaffected, the linear polarization components (Stokes $Q$ and $U$) are significantly different from the magnetized case due to the missing Hanle effect. Even though the optical thickness of the plasma structure is on the order of 1 (see Eq.~\ref{eq:tau}), the symmetry breaking due to the 3D radiative transfer within the medium is sufficiently strong to produce non-zero $U$ signals of the same order as the Stokes $Q$ ones. The inversion of this data using the pixel-by-pixel approach would lead to erroneous conclusions about the presence of magnetic fields.

\subsection{Basis functions and the inversion setup}\label{ssec:basinv}

The MHD-like $\psi$ quantities to invert are: $\chi_{\rm L},\,\Delta\lambda_{\rm D},\,\Gamma_x,\,\Gamma_y,$ and $\Gamma_z$, namely: the line opacity, the line Doppler width, and the Cartesian magnetic field components,  defined in Appendix~\ref{ssec:app-atom}. While in this particular model, all the quantities are dimensionless by definition, in realistic applications, a suitable normalization should be applied. The model atmosphere is static, hence we are not considering any macroscopic velocity field. Both elastic and inelastic collisions can also be (and are) neglected. In this simple model, the line opacity is equivalent to the density, while the Doppler width is equivalent to the kinetic temperature.

Because we adopted a two-level atom with unpolarized lower state, for this example, the upper-level density matrix, $\rho^K_Q(u),$ and the mean radiation field tensors, $\bar J^K_Q$, are completely equivalent in describing the atomic state. In this example, we chose the $\bar J^K_Q$ components as our atomic-like $\xi$ variables: $\bar J^0_0,\,\bar J^2_0,\,\Re \bar J^2_1,\,\Im \bar J^2_1,\,\Re \bar J^2_2,$ and $\Im \bar J^2_2$, where $\Re$ and $\Im$ stand for the real and imaginary parts, respectively. The components with rank $K=1$ remain identically zero because of the adopted model. We normalized the atomic like variables with the disk-center intensity and, in addition, we scaled the components with $K=2$ with a factor $20$ because we can expect the anisotropy of the radiation field to be on the order of few percent of the mean intensity.

For the function basis to approximate the spatial distribution of the problem quantities, we chose the Chebyshev polynomials of the first kind, using the tetrahedron subset (see the central panel of Fig.~\ref{fig:baf}). The maximum orders of the basis functions for the MHD-like quantities are $p(\chi_{\rm L}) = p(\Delta\lambda_{\rm D}) = 2$ and $p(\Gamma_x) = p(\Gamma_y) = p(\Gamma_z) = 1$ -- which, according to Eq.~\eqref{eq:mp}, entails $\Mbaf(\chi_{\rm L})=\Mbaf(\Delta\lambda_{\rm D})=10$ and $\Mbaf(\Gamma_x) = \Mbaf(\Gamma_y) = \Mbaf(\Gamma_z)=4,$ with the total number of MHD-like variables equal to 32. In this example, we can afford to determine the basis from the a priori knowledge of the parameterization of the model (cf. Eqs.~\ref{eq:chir} to \ref{eq:bz}) but, in general, we would have to use adequate techniques to determine them (see Sect.~\ref{sssec:newalg}).
We empirically found via numerical experimentation that the order of the basis of the atomic-like variables sufficient to reach NLTE consistency in our example is $p(\bar J^K_Q)=3$ -- which, using Eq.~\eqref{eq:mp}, gives us $\MbafXi=20$ and the total number of atomic-like variables is $\NXi\MbafXi=5\cdot 20=120$.

Given the relative amplitudes of the Stokes signals in the observations (see top panels in Fig.~\ref{fig:obs}), we set the weights to $w_I=1$, $w_Q=w_U=20$, and $w_V=200$ in the $\chi^2$ function in Eq.~(\ref{eq:chi2}). We note that these are the values of $w_k$ before their normalization to 1. The weight of the NLTE regularization factor is set to $\lambda=10^4$ with the aim to reduce the residual NLTE error below $10^{-4}$. We set $\gamma=1$ for the $\Lloc$ term, a value sufficient to end up with a locally consistent solution in this example, as we show below.

The model was initialized with random values of the variables and we used $\Npp=3$ pilot points, $\Npx=10$ probing pixels, and $\Nloc=10$. The angular quadrature in the pilot points is the 88-point $L=11$ quadrature of \citet{2020quad1}. Alternatively, it might be preferable to use even more accurate quadratures by \citet{2021quad2}. In this example, we do not consider the randomization of the quadrature orientation described in Sect.~\ref{ssec:quad} because the boundary conditions of Eq.~(\ref{eq:clv}) are such that the problem does not suffer from disk-edge artifacts.

We have implemented the meshfree stochastic method in a parallel C code and solved the inversion problem using the ADAM algorithm with parameters $\alpha=0.001$, $\beta_1=0.9$, $\beta_2=0.999$, and $\epsilon=10^{-8}$ \citep[see][for details]{2014arXiv1412.6980K}.

\subsection{Results}

In Fig.~\ref{fig:conv}, we show the convergence of the total loss function, $\LL$, as well as of the penalty functions, $\Llam$ and $\Lloc$, and of the $\chi^2$ function. The $\chi^2$ eventually converges almost to 1 and $\Llam$ and $\Lloc$ decrease to the values corresponding to small deviations from the full consistency. It is worth to say that we have not used any quantitative stopping criteria nor did we extensively experimented with the hyperparameters affecting the convergence. The results in Fig.~\ref{fig:conv} are given only to demonstrate the general behavior of the convergence in the numerical experiments we performed and we note that the problem requires a more thorough analysis.

In terms of CPU time, every iteration shown in Fig.~\ref{fig:conv} requires about 5 seconds using a common contemporary CPU. The total computing time of this inversion with 16\,000 iterations is of about 22 hours. In other words, this is equivalent to about 20 seconds per pixel. It is interesting to compare this time with an estimation of the inversion time of the very same model using the standard $\chi^2$ minimization of a grid-based method. In the PORTA code, that could be used as a $\Lambda$-iteration engine, the CPU time per mesh point, per wavelength, and per quadrature propagation direction is of about $t=10^{-6}$ seconds using the same CPU. One $\Lambda$ iteration with four Stokes parameters therefore requires about $4 N^3 N_\lambda N_\Omega t \approx 4\,300$\,seconds. Assuming, being very optimistic, that we would only need one $\Lambda$ iteration to calculate a derivative of $\chi^2$ with respect to each of the $\psi$ variables and if we neglect the computing time required to solve the NLTE problem at the end of every iteration and the time needed to synthesize the emergent radiation in every step of the $\chi^2$ gradient calculation, we would need about $32\cdot 4\,300\approx 38$\,hours per iteration because the number of the MHD-like variables is 32. Since a realistic number of such iterations is at least $\Nit=100$, the inversion time using such method would require about 3\,800 hours, that is, about a factor 170 more than our solution.\footnote{However, in practice, the number of $\Lambda$ operator evaluations would be larger than one per gradient component in the grid-based methods, hence the speedup of the solution in the meshfree method would be $\gtrsim 340$.} We note that this significant difference is not just due to the fact that our model is quite smooth because in both approaches the CPU time scales with the level of smoothness quantified by $\Mbaf$.

The scatter plots in Fig.~\ref{fig:err} show the correlation of the synthetic and the inverted model quantities at the same spatial points. The diagonal red lines in every panel show the span of the correct values and, in the case of a perfect inversion, all the black points should be on these diognals. Even though the match is not perfect due to the limited accuracy of the basis, the presence of noise, and the slightly premature stopping of the iterative process, the agreement seems to be quite satisfying for all five quantities in the whole 3D domain. In the right panel of Fig.~\ref{fig:mg} we visualize the inferred magnetic field vector, which is indistinguishable at the same plot level as the one in the original model (left panel of the same figure).

The bottom row in Fig.~\ref{fig:obs} shows the emergent radiation corresponding to the inferred model, and compared with the top row in the same figure, we can see that the agreement is remarkable. An example of the good quality of the fit to the observed Stokes profiles at a particular pixel in the field of view is shown in Fig.~\ref{fig:profs}.

\section{Discussion and conclusions\label{sec:concl}}

The unsolved problem of 3DNIP is generally considered to be one of the greatest challenges to face  in the theory of radiation transfer. In this paper, we present a first attempt to solve it. Our approach is not to generalize in a brute-force manner the standard 1D methods into the 3D geometry because such an approach is doomed to failure for a number of reasons.

In inversion methods based on the pixel-by-pixel approach, the NLTE radiative coupling of different regions (in the direction perpendicular to the LOS) of the plasma is usually considered to be a complication. We have developed a meshfree approach that takes this coupling as an advantageous natural regularization which leads to more robust, more physically correct, and computationally faster solutions.

The new method employs the idea that 3DNIP can be solved as an unconstrained minimization problem in which unphysical solutions are allowed during the iterative process. We show that the method has the potential to be much faster than methods based on 3D grids with the recurring evaluations of the $\Lambda$ operator. This approach promises to provide fast solutions, especially in case of relatively smooth models, but it can also provide at least approximate solutions in complex models that would be completely unsolvable using grid-based techniques.

We can summarize the main advantages of the proposed framework as follows:
\begin{enumerate}
\item Consistent 3D\,NLTE solution with scattering polarization, Hanle, and Zeeman effects fully taken into account.
\item Additional conditions for physical consistency, such as $\nabla\cdot\vec B=0$ or the equation of continuity, can be naturally incorporated as penalty terms to the loss function.
\item Since all the pixels are inverted together, the solution is more robust and less sensitive to noise in the data than the pixel-by-pixel methods.
\item Stochastic angular quadratures make it possible to avoid unphysical discontinuities in the boundary illumination in the case of the prominence or filament geometry.
\item Long characteristics lead to more accurate calculation of the radiation field in the pilot points than short characteristics in the grid-based models.
\item There is no numerical error due to the finite spacing of the 3D mesh.
\item The method based on the modern algorithms of stochastic gradient descent is less prone to end up in a local minimum of the loss function.
\item The method can be trivially parallelized for massive HPC facilities. On the other hand, even a single desktop computer can be used to infer at least some information about the global structuring of the plasma properties --- something that would not be possible with grid-based methods.
\item Much smaller memory requirements than in the grid-based methods. There is no need for domain decomposition in order to reduce the memory demands.
\item Given the lack of a grid, the implementation of the code is easier than in grid-based methods.
\end{enumerate}

We go on to list the following disadvantages of the method:
\begin{enumerate}
\item The method is less suitable for problems with a large number of  spectral lines or atomic levels because the number of variables to be inferred could grow and slow down the calculation (see Eq.~\ref{eq:nvar}).
\item In the case of abrupt changes of physical parameters, the parameterized model can have difficulties to accurately describe such discontinuities. This problem is inherent not only to the method presented here, but to every method using an expansion of the physical parameters into a predefined basis.
\item Given the representation of the atomic-like quantities in terms of an expansion in a finite set of basis functions, there may be certain residual NLTE inconsistency in the solution. However, this inconsistency can be reduced to any desired level of precision.
\end{enumerate}

We intentionally do not discuss a number of important issues related to the 3D inversion problem. Among these issues, we have the problems of possible ambiguities, as well as of the local minima, location of the observed structures along the LOS, the convergence criteria, various possibilities for adaptivity of the inversion algorithm, and many others. Last but not least, applications with realistic spectral lines need to be studied. 

\begin{acknowledgements}
J.\v{S}. acknowledges the financial support of the grants \mbox{19-20632S} and \mbox{19-16890S} of the Czech Grant Foundation (GA\v{C}R) and the support from project \mbox{RVO:67985815} of the Astronomical Institute of the Czech Academy of Sciences. We acknowledge the funding received from the European Research Council (ERC) under the European Union's Horizon 2020 research and innovation programme (ERC Advanced Grant agreement No~742265). We are grateful to Roberto Casini and Luca Belluzzi for their insightful comments which helped us to improve the content of the paper.

\end{acknowledgements}

\bibliographystyle{aa}
\bibliography{ms.bib}

\appendix

\section{Implementation details\label{sec:app-impl}}

\subsection{Expansion in the basis functions}\label{ssec:bases}

\begin{figure*}
\begin{center}
\includegraphics[width=0.65\textwidth]{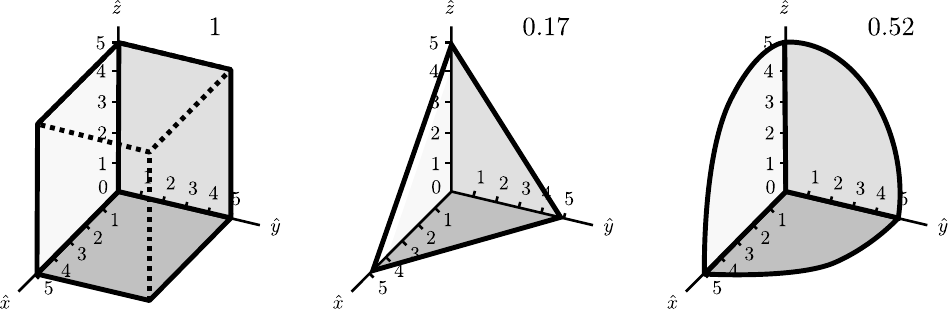}
\end{center}
\caption{Schematic representation of three bases of the same order $p=5$ for approximating the spatial variation of the problem variables in the computational domain. The numbers in the top right part of each diagram indicate the fraction of the number of the considered basis functions with respect to the full basis in the left diagram.}
\label{fig:baf}
\end{figure*}

The choice of the orthonormal basis functions $\{\phi_i(\vec r)\}_{i=1}^M$, with $M$ standing for either $\Mbaf$ or $\MbafXi$, is arbitrary in the sense that any such set with $M=N^3$ is capable of equivalently describing the spatial variation of the physical quantities in a grid-based model with $N^3$ points. However, we need to choose a subset of the basis functions with $M\ll N^3$ that will allow to accurately describe such spatial variation. Therefore, different sets of functions and coordinate systems will provide different results.

One of the possible choices for a 3D basis is to construct it as a product of three one-dimensional (1D) basis functions,
\begin{equation}
\phi_i(\vec r) \to \phi_{klm}(\vec r) =\phi_k(x)\phi_l(y)\phi_m(z)\;,
\end{equation}
where the functions $\phi_n$ are taken from a suitable 1D basis. For instance, we can use $\phi_k(x)=T_k(x)$ with $T_k(x)$ being a Chebyshev polynomial of the first kind and order $k$ \citep{abramowitz14}. Alternatively, we could use the basis of a discrete cosine transform or any other set of orthogonal functions.

It is natural to restrict the orders $k$, $l$, and $m$ to be smaller or equal to a certain integer $p$. This would typically imply a restriction to the $(p+1)^3$ smoothest functions (see the left panel of Fig.~\ref{fig:baf} for $p=5$). Since the computing time is always a concern, we can even use a smaller subset of this basis. In our numerical experiments, we have tested the shapes of the bases shown in the middle and right panels of Fig.~\ref{fig:baf}. Apart from the $(p+1)^3$ basis in the left panel, two alternative bases with the same order of functions but smaller $M$ are the bases satisfying the conditions $k+l+m\le p$ with 
\begin{equation}
M=(p+1)(p+2)(p+3)/6
\label{eq:mp}
\end{equation}
(middle panel of Fig.~\ref{fig:baf}) and $k^2+l^2+m^2\le p^2$ with
\begin{equation}
M\approx \pi (p+1)^3/6
\end{equation}
(right panel in Fig.~\ref{fig:baf}). Using such subsets can lead to a significant savings in CPU time, while the approximation capabilities may  be only slightly deteriorated. The choice of the subset is indeed problem-dependent and we will discuss it in future publications. At this point, we only note that it is not necessary to use the same subset of basis functions for every axis. For instance, reducing the $\hat z$ basis to $\phi_0(z)=1$ makes the problem effectively 2.5D with physical quantities being constant along the line of sight (given the geometry in Fig.~\ref{fig:geom}).

Apart from the possibility to construct the 3D basis out of the 1D functions in the Cartesian coordinates, it might be useful to use orthogonal functions in different coordinate systems, such as in spherical or cylindrical coordinates.

\subsection{Normalization of the quantities}\label{app:normal}

In the calculation of the $\chi^2$ function, it is typical to use different weights $\{w_k\}_{k=0}^3$ for different Stokes parameters (see Eq.~\ref{eq:chi2}). This is due to the fact that the Stokes parameters are usually on different orders of magnitude. If the weights of the Stokes parameters were all the same, the inversion algorithm would try to fit the intensity profile, while the sensitivity to polarization components would be very small.

A similar situation can occur with the $\theta$ parameters of the problem variables. For instance, the atomic level population, $\rho^0_0$, is usually much larger than the level alignment, $\rho^2_0$. Since we want the loss function of Eq.~(\ref{eq:llamdef}) to be sensitive to the errors of variables with different order of magnitude, it is desirable to choose the $\theta_i$ variables in such a way that they end up having similar orders of magnitude. This can be achieved by using a proper normalization. The same applies to all the variables in the $\theta$ vector that should be suitably normalized in order to have similar orders of magnitude for their ``typical'' physical values.

\subsection{Local consistency regularization}\label{sec-app:lloc}

The third term on the right-hand side of Eqs.~(\ref{eq:ll2}) and (\ref{eq:llapprox}) stands for a penalty for the deviation of the model parameters from physical consistencies other than the NLTE consistency. These include penalties for negative atomic level population, an unphysical degree of atomic polarization, a violation of the magnetic field zero divergence condition ($\nabla\cdot\vec B=0$) or of the equation of continuity, and any other condition, depending on the particular problem. The evaluation of $\Lloc$ does not involve radiative transfer calculations but, rather, a check of the values and possibly derivatives of quantities in $\Nloc$ points. This calculation is therefore very fast. The $\Lloc$ term is important as an additional physical regularization of the problem that allows us to incorporate relatively complex physical conditions into a small piece of a numerical code.

The loss function of the local physical consistency, $\Lloc$, can be written as an average over $\Nloc$  random points in the following form:
\begin{equation}
\Lloc=\frac{1}{\Nloc}\sum_{i=1}^{\Nloc}\sum_{\alpha}
g_\alpha(\vec r_i; \theta)
\,,
\label{eq:glocgen}
\end{equation}
where $\theta$ stands for the vector of model variables of Eq.~(\ref{eq:thetadef}) and the summation over $\alpha$ stands for different problem-dependent penalties, $g_\alpha$. These non-negative differentiable penalty functions should take zero value for the physically consistent solutions at any point, $\vec r_i$.

This very general formulation can be easily understood considering particular examples. If we use $x$ to represent a particular positive quantity that is constructed out of $\theta$ (such as an atomic-level population, plasma temperature, or any function of the $\theta$ variables), it is useful to define a penalty of the form
\begin{equation}
g_\alpha(x) \equiv g_+(x;t)=
\begin{dcases*}
0 & for $x > 0$ \\
(x/t)^2 & for $x\le 0$
\end{dcases*}\,,
\label{eq:g+}
\end{equation}
where the scaling parameter, $t,$ is to be chosen depending on the variable and it controls the sensitivity of the loss function to this particular quantity. Thus, the smaller the value of $t,$ the stronger the penalties for any violations of the condition. In order to avoid convoluted notation, we do not explicitly indicate the point of $g_+$ evaluation, $\vec r_i$; the quantity $x$ in $g_\alpha$ is always evaluated in all the $\Nloc$ points and the resulting penalties are summed according to Eq.~(\ref{eq:glocgen}). If $x$ is a quantity that must always be larger than some other quantity or a constant, $y$, we can use a penalty of the form
\begin{equation}
g_\alpha(x) \equiv g_>(x,y;t)=
\begin{dcases*}
0 & for $x > y$ \\
\left(\frac{x-y}{t}\right)^2 & for $x \le y$
\end{dcases*}\;.
\end{equation}
Analogously, if a quantity needs to be zero for physical reasons, the function:
\begin{equation}
g_\alpha(x) \equiv g_0(x;t)=(x/t)^2
\label{eq:gzero}
\end{equation}
provides a way to penalize deviations from the zero value. An example of such quantity is the magnetic field divergence, $x=\nabla\cdot\vec B=\partial_x B_x+\partial_y B_y+\partial_z B_z$. Additional local physical constrains can easily be incorporated into $\Lloc$ (i.e., the equation of continuity in the stationary models, $x=\nabla\cdot(\rho\vec v) $ together with the $g_0$ penalty, the condition of magneto-hydrostatic equilibrium, etc.)\footnote{We use the same symbol $t$ in Eqs.~(\ref{eq:g+})--(\ref{eq:gzero}) in order to avoid cluttered notation but its value indeed depends on the context and the order of magnitude of $x$.}

\section{Example model details}\label{sec:app-model}

This appendix describes some details of the academic model discussed in Sect.~\ref{sec:example}.

\subsection{Model atom and the Hanle and Zeeman effects\label{ssec:app-atom}}

Our example model atom is a normal Zeeman triplet with angular momenta $J_l=0$ and $J_u=1$ for the lower and upper levels, respectively. Given that the collisions in our model are considered negligible, the atomic density matrix of the upper level at a given spatial point, $\rho^K_Q(u)$, is fully determined by the local mean radiation field tensor, $\bar J^K_Q$, and the magnetic field $\vec B$. Instead of $\vec B$ in the units of G, we use the dimensionless vector
\begin{equation}
\vec\Gamma = 8.79\cdot 10^6 g_u \frac{\vec B}{A_{ul}}\;,
\end{equation}
where $g_u$ is the upper-level's Land\'e factor and $A_{ul}$ is the Einstein coefficient of spontaneous emission in $s^{-1}$ \citep[see][for details]{LL04}. It follows that $\Gamma=|\vec\Gamma|$ is the factor commonly used to quantify the Hanle effect \citep[see, e.g., the discussion after Eq.~5 of][where $\Gamma_u$ is used instead of $\Gamma$]{2018ApJ...863..164D}. For $\Gamma=1$, the spectral line is most sensitive to the Hanle effect, and we call the corresponding magnetic field the critical Hanle field.

The irreducible components of the line source function are given by Eq.~(5) of \citet{2018ApJ...863..164D} where $w^{(2)}_{J_uJ_l}=w^{(2)}_{1\,0}=1$, $\epsilon=0$, and $\delta=0$. The matrix elements $M_{ij}$ in that equation are defined in terms of the orientation of the magnetic field vector with respect with the local vertical and can be found in the appendix therein. The source functions of the Stokes parameters then follow from Eq.~(6) of the same paper.

The line absorption profiles are Gaussian with the Doppler width $\Delta\lambda_{\rm D}$:
\begin{equation}
\varphi(\lambda)=\frac 1{\Delta\lambda_D\sqrt\pi}\exp\left(-\frac{\lambda^2}{\Delta\lambda_D^2}\right)
,\end{equation}
where the wavelength $\lambda$ is defined as the distance to the line center for convenience.

In this example, we also account for the longitudinal Zeeman effect assuming that the magnetic field is sufficiently weak so that the Zeeman splitting is much smaller than the line Doppler width and thus only the Stokes~$V$ signal is affected by the Zeeman effect. The radiative transfer equation for circular polarisation in the weak field limit is then \citep{LL04}:
\begin{eqnarray}
\frac{{\rm d}V(\lambda)}{{\rm d}s}=-\chi_L\varphi(\lambda)V(\lambda)\nonumber\\
-\Gamma \frac{\alpha}{\Delta\lambda_{\rm D}}\cos\theta_{\Gamma}\frac{\lambda}{\Delta\lambda_{\rm D}}\varphi(\lambda)\chi_L\left[I(\lambda)-S_L\right]\,,
\end{eqnarray}
where $\varphi(\lambda)$ is the local absorption profile, whose Gaussian form has been taking into account to derive this formal expression, $\Delta\lambda_{\rm D}$ is the Doppler width, $\chi_{\rm L}$ is the line opacity, $S_{\rm L}$ is the intensity line source function, and $\theta_\Gamma$ is the angle between the propagation direction and the magnetic field vector. The value of $\alpha$ is proportional to the effective Land\'e factor and we set it to $\alpha=0.004$ for our academic case. If $\alpha\ll 1$, the weak field limit is satisfied because we typically have $\Gamma\sim 1$ and $\Delta\lambda_{\rm D}\sim 1$.

\subsection{Spatial variation of the physical quantities\label{sec:app-plasma}}

The plasma in our synthetic example model is a cloud with a spherically symmetric distribution of line opacity and Doppler width. For the sake of demonstrating the method, we define a simple functional form of the relevant quantities:
\begin{align}
 \chi_{\rm L}(\vec r) &= 2(1-r^2)\;, \label{eq:chir}  \\
 \Delta\lambda_{\rm D}(\vec r) &= 1+r^2\;, \label{eq:lambdar}   \\
 \Gamma_x&=1-2x-y\;, \\
 \Gamma_y&=1+x+y\;, \\
 \Gamma_z&=-x+2y+z\;,  \label{eq:bz}
\end{align}
where $r=\sqrt{x^2+y^2+z^2}$ is the radial distance from the center of the $[-1,1]^3$ model. The above magnetic field configuration is shown in Fig.~\ref{fig:mg}.

It follows from Eqs.~(\ref{eq:chir}) and (\ref{eq:lambdar}) that the maximum total optical thickness of the medium is in the domain center ($x=y=0$) and it is equal to
\begin{align}
\tau_{\rm max}&=\int_{-1}^1 {\rm d}z\; \chi_L(z) \varphi(\lambda=0,z)\\
&=\frac 2{\sqrt\pi}\int_{-1}^1 {\rm d}z\;\frac{1-z^2}{1+z^2}=\frac2{\sqrt\pi}(\pi-2)\approx 1.3\;.
\label{eq:tau}
\end{align}

\end{document}